\documentclass[]{aa}

\usepackage[varg]{txfonts}
\usepackage{natbib}
\bibpunct{(}{)}{;}{a}{}{,} 
\usepackage{graphicx}
\usepackage{placeins}
\usepackage{ulem}

\authorrunning{S. Th\"olken et al.}

\title{X-ray analysis of the galaxy group UGC03957 beyond $R_{200}$ with Suzaku}

\author{Sophia Th\"olken\inst{\ref{inst1}}\
\and Lorenzo Lovisari\inst{\ref{inst1},}\inst{\ref{inst2}}
\and Thomas H. Reiprich\inst{\ref{inst1}}
\and Jan Hasenbusch\inst{\ref{inst3}}}

\institute{Argelander-Institut f\"ur Astronomie, Universit\"at Bonn, Auf dem H\"ugel 71, 53121 Bonn, Germany\label{inst1}\\
\email{thoelken@astro.uni-bonn.de} \and
Harvard-Smithsonian Center for Astrophysics, 60 Garden Street, Cambridge, MA 02138, USA\label{inst2} \and
Physikalisches Institut, Universit\"at Bonn, Nussallee 12, 53115 Bonn, Germany\label{inst3}
}

\date{Received date /
Accepted date }

\abstract {In the last few years, the outskirts of galaxy clusters have been studied in detail and the analyses have brought up interesting results such as indications of possible gas clumping and the breakdown of hydrostatic, thermal, and ionization equilibrium. These phenomena affect the entropy profiles of clusters, which often show deviations from the self-similar prediction around $R_{200}$. However, significant uncertainties remain for groups of galaxies. In particular the question, of whether entropy profiles are similar to those of galaxy clusters.} {We investigated the gas properties of the galaxy group UGC03957 up to $1.4\,R_{200} \approx 1.4\,$Mpc in four azimuthal directions with the Suzaku satellite. We checked for azimuthal symmetry and obtained temperature, entropy, density, and gas mass profiles. Previous studies point to deviations from equilibrium states at the outskirts of groups and clusters and so we studied the hydrodynamical status of the gas at these large radii.} {We performed a spectral analysis of five Suzaku observations of UGC03957 with ${\sim} 138$\,ks good exposure time in total and five Chandra snapshot observations for point source detection.  
We investigated systematic effects such as point spread function and uncertainties in the different background components, and performed a deprojection of the density and temperature profile.} {We found a temperature drop of a factor of ${\sim} 3$ from the center to the outskirts that is consistent with previous results for galaxy clusters. The metal abundance profile shows a flat behavior towards large radii, which is a hint for galactic winds as the primary ICM enrichment process. The entropy profile is consistent with numerical simulations after applying a gas mass fraction correction. {Feedback processes and AGN activity might be one explanation for entropy modification, imprinting out to larger radii in galaxy groups than in galaxy clusters}. Previous analyses for clusters and groups often showed an entropy flattening or even a drop around ${\sim} R_{200}$, which can be an indication of clumping or non-equilibrium states in the outskirts. Such {entropy behavior} is absent in UGC03957. The gas mass fraction is well below the cosmic mean but rises above this value beyond $R_{200}$, which could be a hint for deviations from hydrostatic equilibrium at these large radii. By measuring the abundance of the $\alpha$-elements Si and S at intermediate radii we determined the relative number of different supernovae types and found that the abundance pattern can be described by a relative contribution of $80$\% -- $100$\% of core-collapse supernovae. This result is in agreement with previous measurements for galaxy groups.} {}

\keywords{galaxies: groups: general - galaxies: groups: individual: UGC03957 - X-rays: galaxies: clusters}

\begin{document}

\maketitle

\section{Introduction}\label{sec:introduction}

\begin{figure}[htbp]
\resizebox{\hsize}{!}{\includegraphics{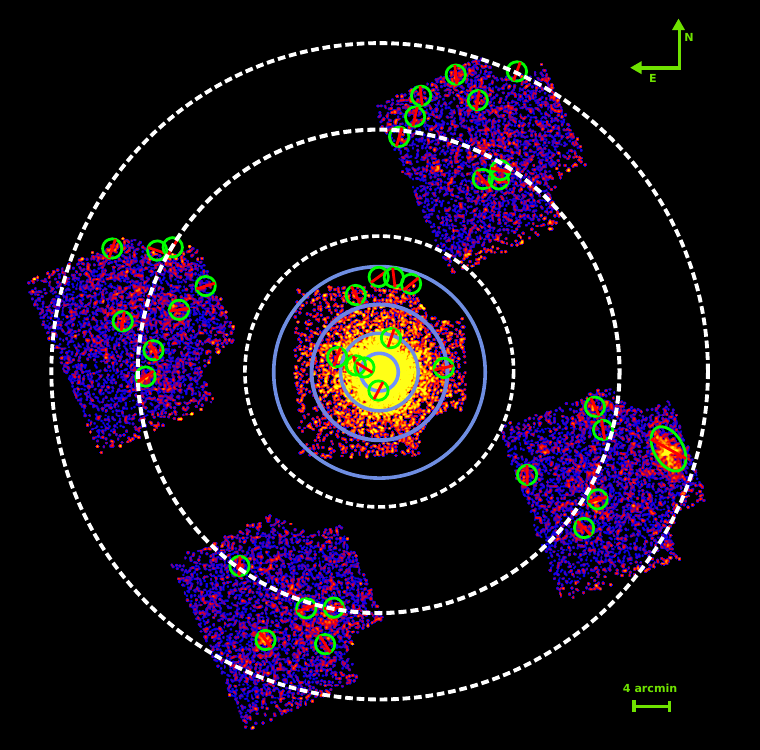}}
\caption{Exposure corrected mosaic image of UGC03957. The central observation was performed in 2006; the outer observations (north, east, south and west) are from 2012. For observation details see Tab. \ref{tab:observations}. Light blue solid regions correspond to $0\arcmin$ -- $2\arcmin$,$2\arcmin$ -- $4\arcmin$,$4\arcmin$ -- $7\arcmin$, and $7\arcmin$ -- $11\arcmin$ (labeled from one to four outwards in the following), white dashed regions to $14\arcmin$ -- $25\arcmin$ and $25\arcmin$ -- $34\arcmin$ (labeled five and six in the following). All removed point sources detected with Chandra are indicated by green circles. The image is not corrected for vignetting and only used for demonstration purposes.}
\label{fig:pointings}
\end{figure}

The outskirts of galaxy clusters, and in particular galaxy groups, still remain unexplored to a large extent.  X-ray studies are the best tool for investigating temperature and metallicity of the hot gas component in these objects, which accounts for ${\sim} 90\%$ of the total baryon content. In the recent years, several analyses investigating the outskirts have been performed (e.g., \citealp{2014PASJ...66...85S}, for a review see \citealp{2013SSRv..tmp...33R}). However, most of them deal with properties of galaxy clusters and there is a clear lack of detailed analyses at the low mass end. These analyses have brought up interesting results such as possible non-equilibrium states (e.g., {\citealp{2010PASJ...62..371H}}, \citealp{2011PASJ...63S1019A}, \citealp{2012PASJ...64...49A}) or gas clumping (\citealp{2011Sci...331.1576S}) due to the infalling material from the large-scale structure around the virial radius. For understanding these effects and their interplay in detail, galaxy groups can give important insights. In these objects the non-gravitational processes are expected to be more important owing to the shallower gravitational potential well. In a self-similar picture of the cluster formation process, galaxy groups should behave as scaled down versions of galaxy clusters regarding, e.g., temperature, density, and entropy profile. In particular the entropy profile is an important indicator of the hydrodynamical status of the gas. However, for galaxy clusters several previous studies (e.g., \citet{2013ApJ...766...90I}, \citet{2012MNRAS.424.1826W} and \citet{2012MNRAS.427L..45W}) have found a flattening or even a drop of the entropy profile at large radii compared to the expectation from numerical simulations of the gravitational collapse formation model performed by \citet{2005RvMP...77..207V}. This behavior may indicate a breakdown of thermal equilibrium between electrons and protons (e.g., \citealp{2011PASJ...63S1019A}) or inhomogeneous gas distributions in the outskirts; the latter possibility is also supported by simulations performed by \citet{2011ApJ...731L..10N}. A study of a galaxy group by \citet{2013ApJ...775...89S} have also found an entropy drop, whereas \citet{2012ApJ...748...11H} have obtained an entropy profile for a fossil group in agreement with the simulations by \citet{2005RvMP...77..207V}. {Other studies of entropy profiles for clusters and groups have been performed by, e.g., \citet{2012ApJ...759...87C} and \citet{2015ApJ...805..104S} yielding different behaviors of the entropy profiles at large radii.}
Therefore, the question remains whether galaxy groups behave in a self-similar way compared to galaxy clusters regarding, e.g., the entropy profile.

Self-similarity is an important assumption when dealing with scaling relations, in particular at the low mass end of galaxy groups.  
As measured by, e.g., \citet{2011A&A...535A.105E} and \citet{2015A&A...573A.118L} scaling relations often show deviations from the self-similar prediction in this regime. However, the scatter is still large and more detailed studies are required out to the virial radius to avoid biases due to the extrapolation of the measured profiles. \citet{2012MNRAS.421.1583M} among others have studied the $L_X$ - $T$ relation for 114 clusters. They combined their cluster sample with the cool core cluster sample of \citet{2009A&A...498..361P} to reach the low mass regime and temperatures $<$3.5\,keV. In this regime they found a strong deviation from the self-similar prediction. One possibility for a deviating process in clusters and groups is AGN feedback (e.g., \citealp{2014A&A...572A..46B}). AGN heating might have a significant impact at larger radii in galaxy groups than in galaxy clusters because of their lower mass, which leads to further expansion of the heated gas. Other non-gravitational processes such as galactic winds or star formation can also play a significant role in low mass systems, while they should be less important in galaxy clusters. \citet{2013A&A...551A..22E} investigated the average entropy profile of 18 galaxy clusters confirming an entropy excess at smaller radii and a better agreement with the numerical simulations farther out. This entropy excess suggests that non-gravitational effects such as feedback from the central AGN or preheating processes are present in the intracluster medium (ICM).

Another aspect of the evolution of galaxy clusters and groups is the chemical enrichment history. Measuring the abundance and especially individual abundances of $\alpha$-elements can give important insights into the chemical evolution of the ICM. {This has been done previously by, e.g., \citet{2004A&A...420..135T}, \citet{2007A&A...465..345D}, \citet{2007PASJ...59..299S}, \citet{2007PASJ...59S.327M}, \citet{2008PASJ...60S.317T}, \citet{2009PASJ...61S.337K}, and \citet{2009A&A...493..409S}}. The heavy elements that can be found in the ICM are thrown out by supernova explosions into the surrounding medium. This material is then distributed to the ICM, mainly by galactic winds and ram pressure stripping. As was simulated by \cite{2007A&A...466..813K}, clusters primarily enriched by ram pressure stripping show a steeper abundance profile than clusters where the enrichment is dominated by galactic winds, i.e., ram pressure stripping acts more efficiently in the dense cluster centers whereas galactic winds are present at all radii. The radial profile is not the only important aspect, however; the ratio between different elements also contains information about the past. 
The ratio of alpha-elements to iron abundances gives information about the amount of Supernovae Type Ia (SNIa) compared to core-collapse supernovae (SNCC) events in the past (e.g., \citealp{2015A&A...575A..37M}, \citealp{2015ApJ...811L..25S}, \citealp{2011A&A...528A..60L}). This ratio can be computed for different supernovae yield models and in principle allows to distinguish between the models (e.g., \citealp{2007ApJ...667L..41S}). 

Measuring all the mentioned profiles and properties of clusters and in particular of galaxy groups is challenging as the surface brightness (SB) drops quickly towards the outskirts and therefore the treatment of the background emission is crucial. The Suzaku satellite is of special importance for these kinds of analyses because of its low and stable instrumental background due to its low Earth orbit and short focal length. Here we present an X-ray analysis of the galaxy group UGC03957 with Suzaku reaching $1.4R_{200}$, where $R_{200}=23.7\arcmin$ is obtained from the Suzaku data in this work (see Sec. \ref{sec:discussion_mass}). We measure temperature, metallicity, density, entropy, surface brightness, and gas mass fraction profiles up to and beyond $R_{200}$ and the entropy profile. In addition we investigate the ratio of SNIa to SNCC from the abundance pattern of $\alpha$-elements in the center and compare different SNIa yield models. Throughout the analyses we assume a flat universe with $H_0=70$\,km\,s$^{-1}$\,Mpc$^{-1}$ and $\Omega_{\Lambda} = 0.73$. All errors are given at a 68\% confidence level.

\begin{figure*}[htbp]
\includegraphics[scale=0.3]{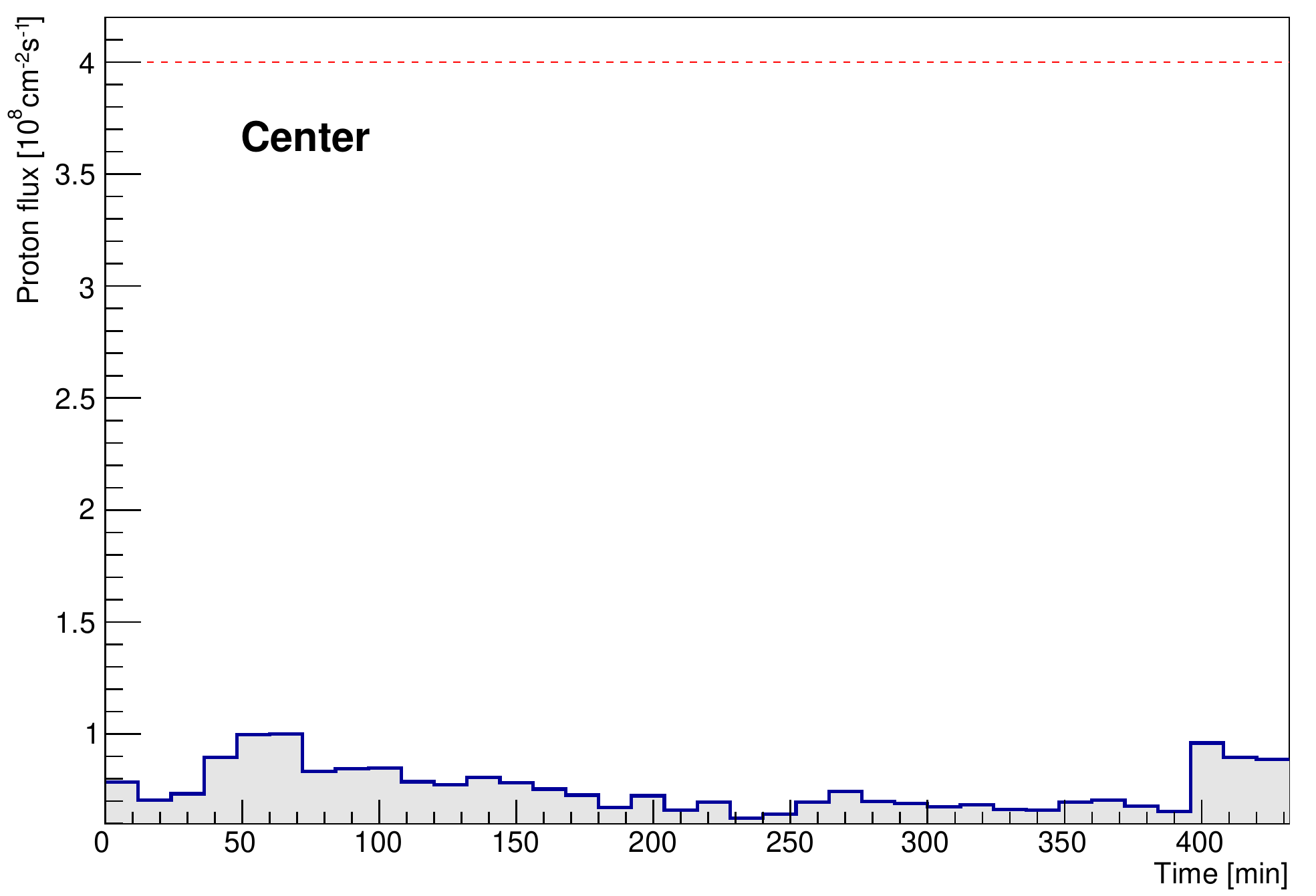}
\includegraphics[scale=0.3]{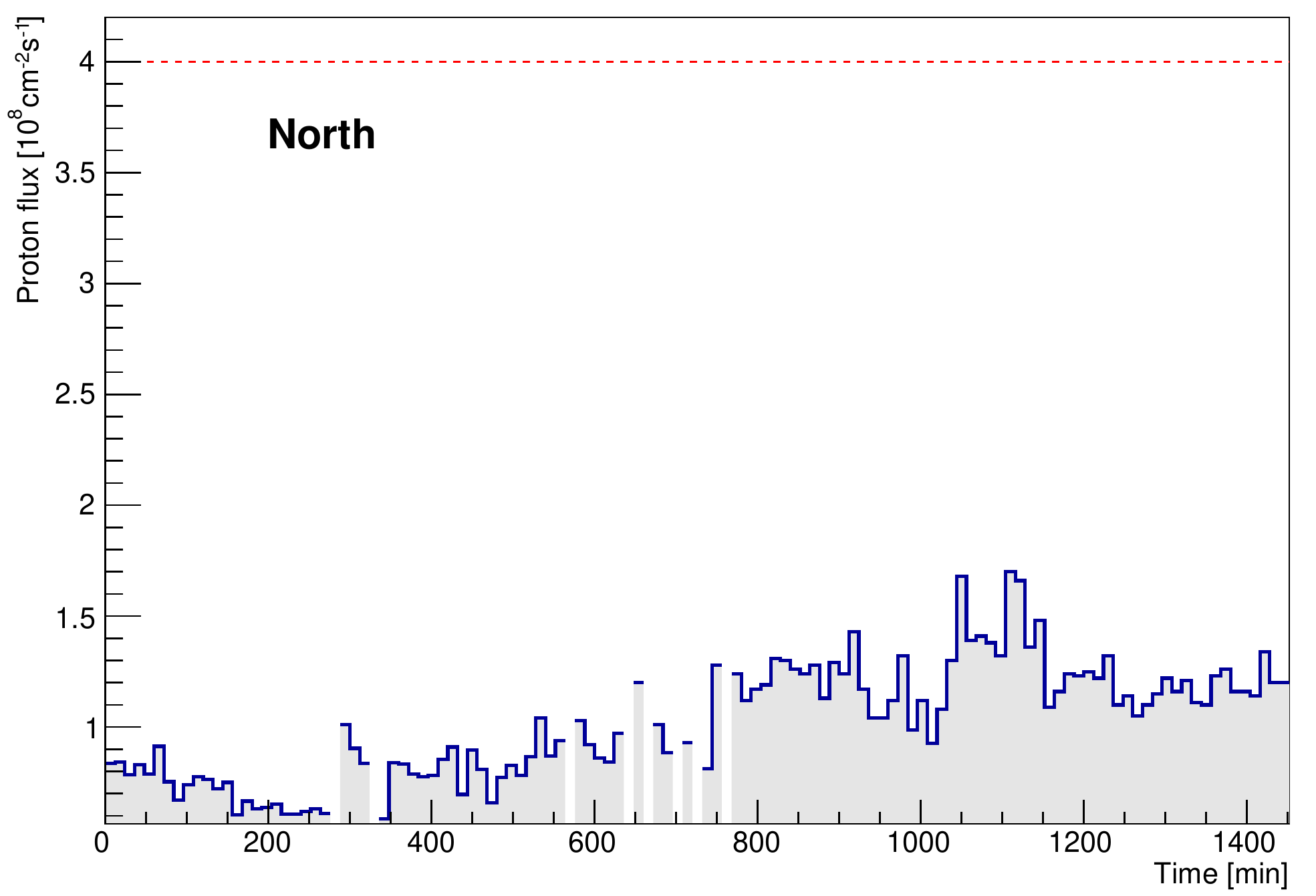}
\includegraphics[scale=0.3]{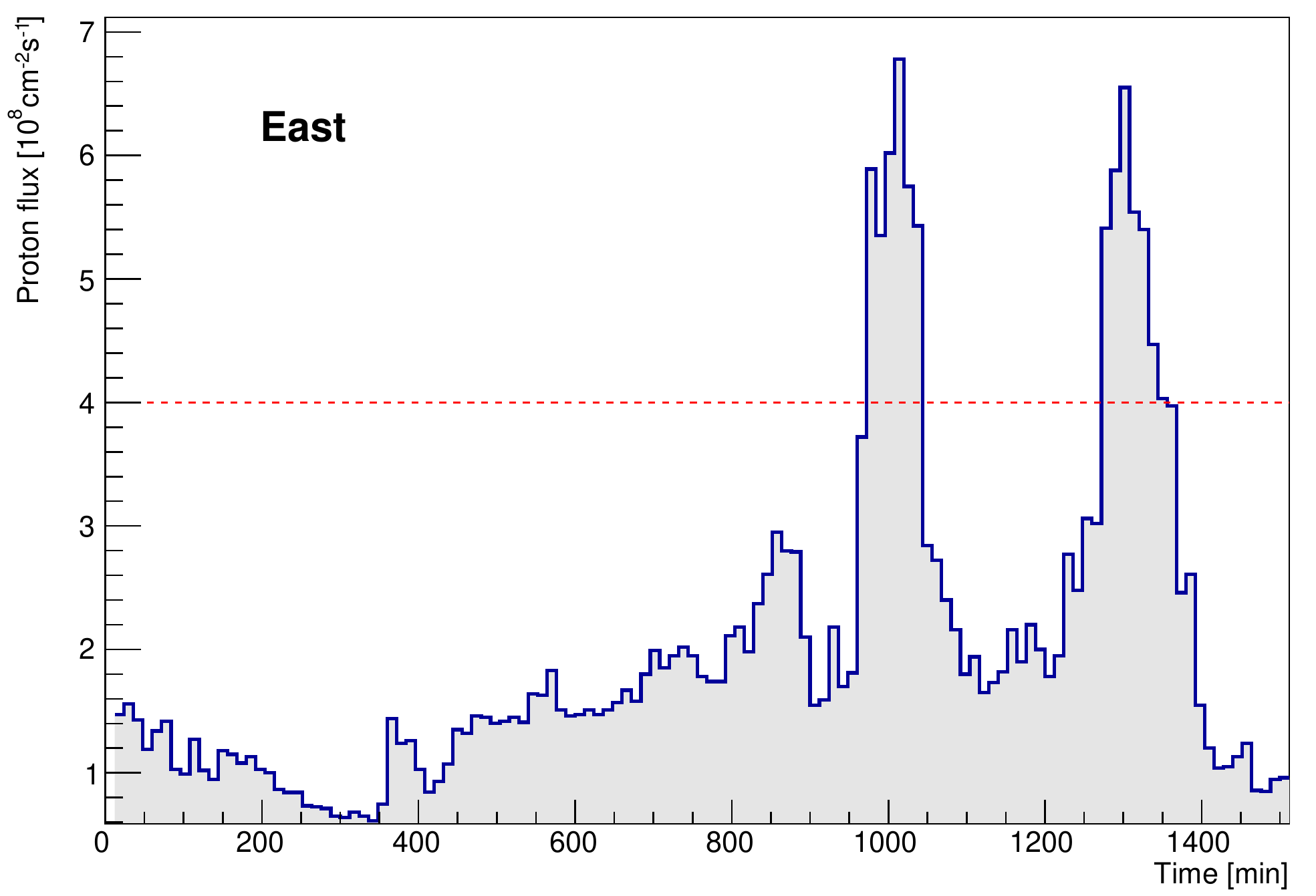}
\includegraphics[scale=0.3]{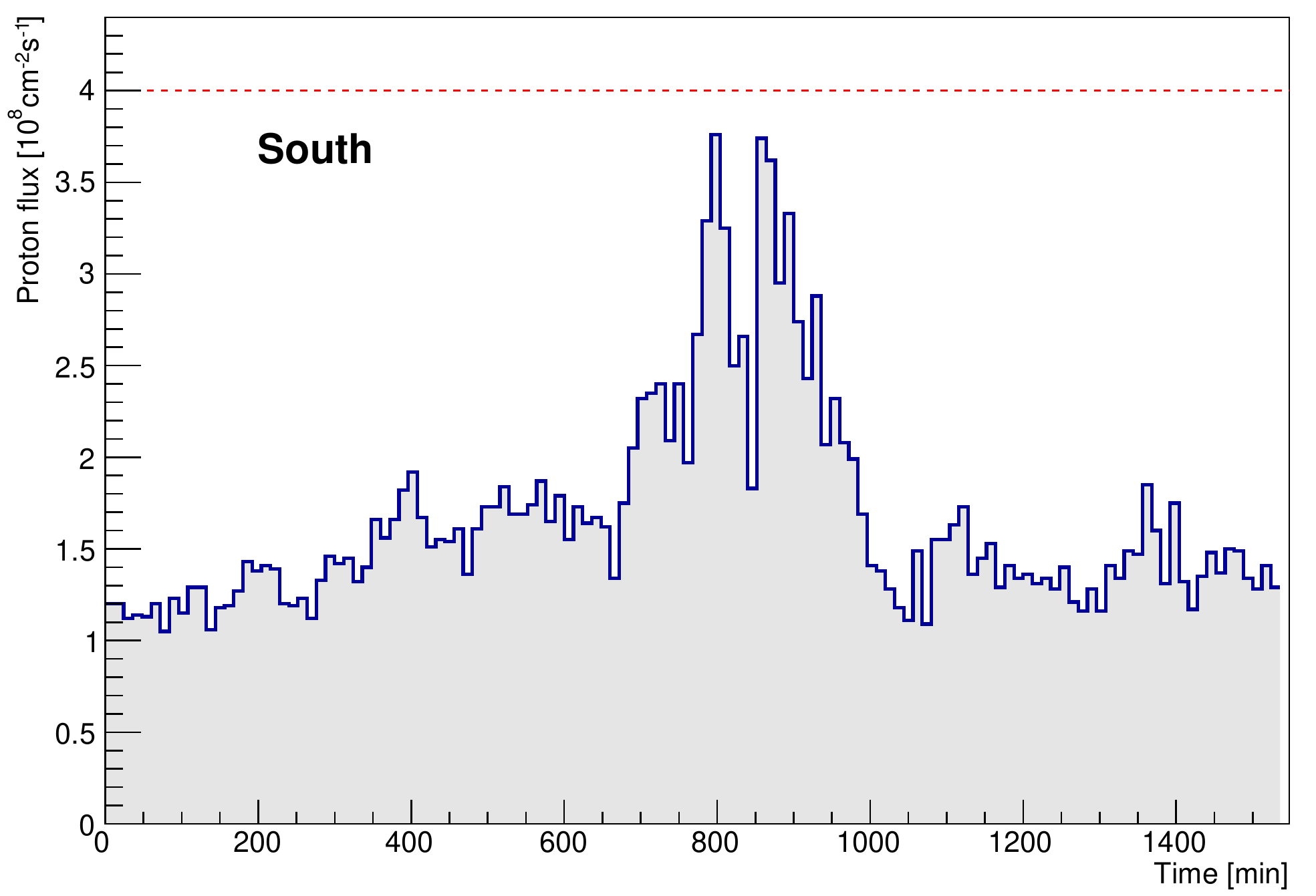}
\includegraphics[scale=0.3]{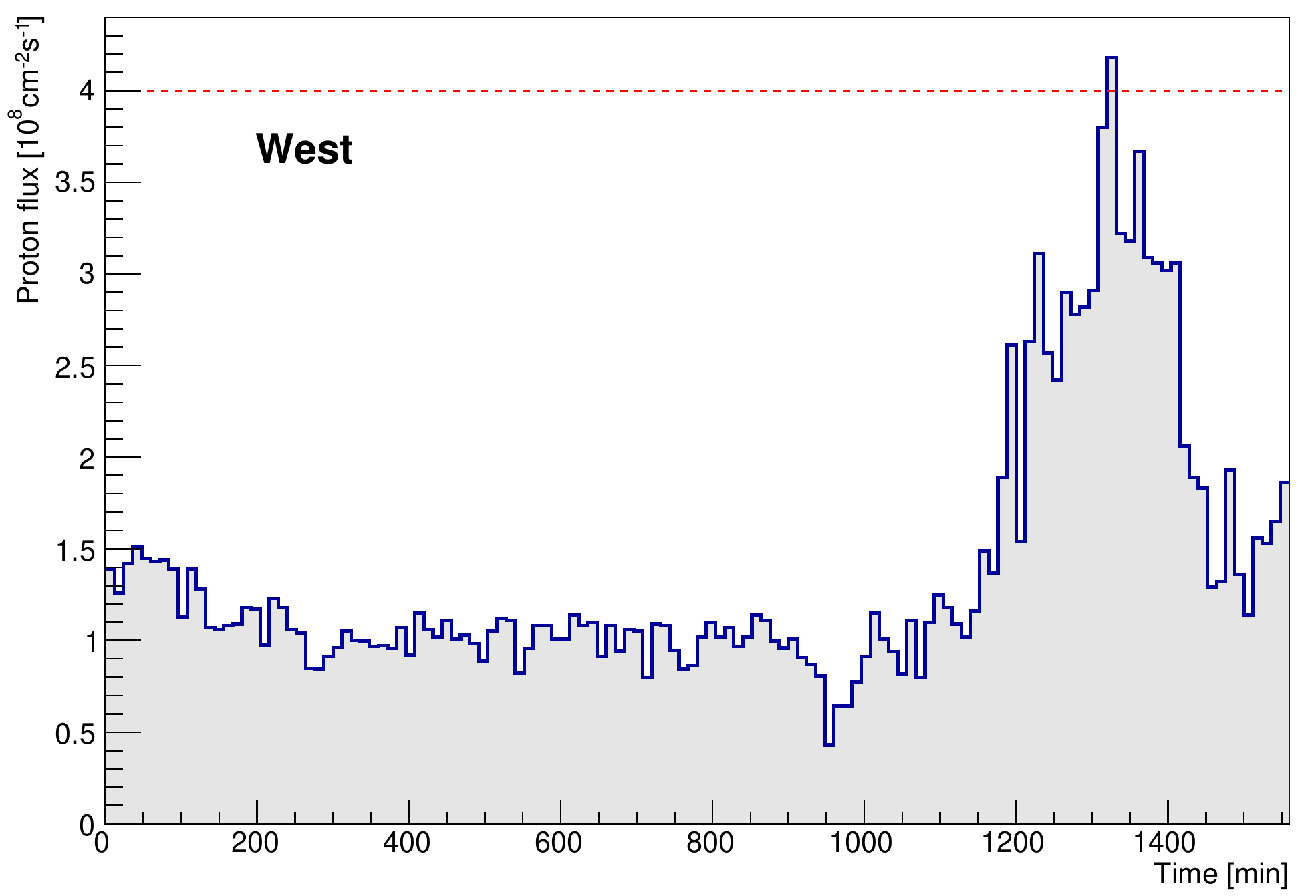}
\caption{Proton flux during the five Suzaku observation time periods measured with SWEPAM/SWICS and corrected for the travel time of the solar protons to Earth. If no data is available, the bin is left empty. The red dashed line shows the limit for flaring determined by \citet{2007PASJ...59S.133F}. }
\label{fig:proton_flux}
\end{figure*}

\section{Observations and data reduction}\label{sec:observations}
\subsection{Suzaku}
The galaxy group UGC03957 is one of the most luminous local groups with a redshift of $z=0.034$. We analyze five Suzaku observations performed with the XIS instrument aboard of Suzaku with 138\,ks total cleaned exposure time (see Tab. \ref{tab:observations}). One short archival observation (analyzed as part of a sample in \citealp{2009ApJ...690..879S}) points towards the center of the group, while we placed four additional deeper observations in each azimuthal direction around the center (called north, east, south, and west observations in the following) as shown in Fig. \ref{fig:pointings}. These observations are very well suited to study possible azimuthal deviations in the outskirts, each of the four reaching beyond $R_{200}$. The central Suzaku observation was taken in 2006, while the other four were performed in March 2012. XIS2 was damaged by a micrometeorite hit in November 2006; therefore, only XIS0, XIS1, and XIS3 data can be used for the analyses of the outer observations, while data of all four XIS chips is available for the central observation. Further details are listed in Tab. \ref{tab:observations}.
A first estimate for the radius $R_{200}$ where the gas density is 200 times the critical density of the universe was determined using Rosat All-Sky Survey (RASS) data (\citealp{2002ApJ...567..716R}). This yields a value of $R_{200}=34.06\arcmin$, which is also the maximum radius we reach with our observations. We determine $R_{200}$ using Suzaku data in Sec. \ref{sec:discussion_mass}.

\begin{table}
\caption{Details of all analyzed observations of UGC03957. The exposure time is given after data reduction.}             
\centering                                      
\begin{tabular}{c c c c c c c}          
\hline \hline                      
 & Date & (R.A., Dec.) & Exp. Time & Obs-ID \\ \hline
center & 2006 Apr & (115.238, 55.407) &~~9.5\,ks  & 801072010 \\ 
north & 2012 Mar & (114.899, 55.790) & 28.2\,ks & 806091010 \\ 
east & 2012 Mar& (115.959, 55.476) & 32.7\,ks  & 806094010 \\ 
south & 2012 Mar& (115.511, 55.004) & 34.0\,ks  & 806092010 \\ 
west & 2012 Mar& (114.537, 55.221) & 33.9\,ks & 806093010 \\
\hline \hline                                             
\end{tabular}
\label{tab:observations}
\end{table}

The data reduction was performed using CALDB version 20150105 and followed the standard reduction procedure as described in the Suzaku data reduction guide. This includes the tasks {\it xiscoord} to calculate event coordinates, {\it xisputpixelquality} to assign the quality code to each event (e.g., falls in bad pixel), {\it xispi} to calculate pulse invariant values using gain- and charge-transfer-inefficiency correction, {\it xistime} to assign correct arrival times, and {\it cleansis} to identify anomalous pixels. The requirement on the geomagnetic cutoff rigidity is COR2$>$6 and we removed events falling in the second trailing rows of the charge injection rows.
We selected six annular regions around the center as shown in Fig. \ref{fig:pointings}: the four inner regions ($0\arcmin$ -- $2\arcmin$, $2\arcmin$ -- $4\arcmin$, $4\arcmin$ -- $7\arcmin$, and $7\arcmin$ -- $11\arcmin$) covering the central Suzaku observation and the two outer regions ($14\arcmin$ -- $25\arcmin$ and $25\arcmin$ -- $34\arcmin$) covering the four observations of the outskirts.
Owing to the long time period between the central and the outskirts observations of almost six years, the central observation was analyzed separately from the four outer observations. 

To investigate the impact of flares during the observations that may be caused by high solar-wind-charge-exchange emission, we checked the solar proton flux using SWEPAM/SWICS Level 3 data\footnote{\texttt{ http://www.srl.caltech.edu/ACE/ASC/}}. This data includes measurements of the proton speed and proton density from both instruments, but SWEPAM data is preferentially used owing to the higher quality, and only gaps in this data set are filled with SWICS data. For slow solar winds SWEPAM may underestimate the proton flux, and SWICS data is used instead. The measured fluxes for all pointings are shown in Fig. \ref{fig:proton_flux}. The travel time of the solar protons to Earth was considered.
 We found a very low proton flux during the time period of the central observation so that the impact of flares for this observation is negligible. During the observation time period of the outskirts, the proton flux is higher and in the case of the east, south, and west observations reaches the limit of ${\sim}4\times10^{8}$\,cm$^{-2}$\,s$^{-1}$ (as was determined by \citealp{2007PASJ...59S.133F}), which can lead to flare contamination of the lightcurves. Therefore, we applied a three-sigma clipping to the lightcurves for all outer observations and filtered the corresponding time intervals.

\subsection{Chandra}
In 2013 and 2014, we obtained four supporting Chandra snapshot observations of ${\sim} 10$\,ks exposure each to detect point sources in the north, east, south, and west Suzaku pointings. For point source detection in the central observation we used archival Chandra data from 2006 of ${\sim} 8$\,ks exposure.
The Chandra data reduction was performed using the CIAO software (CIAO 4.5, CALDB 4.6.7). The data was reprocessed from the “level 1” events files  using the contributed script {\it chandra\_repro}. Periods of high background were cleaned by creating  a  lightcurve  with  the  suggested values from  Markevitch’s Cookbook\footnote{\texttt{http://cxc.harvard.edu/contrib/maxim/acisbg/COOKBOOK}} and the {\it lc\_clean} algorithm. Point sources were identified using the {\it wavdetect} algorithm using a range of wavelet radii between 1 and 16 pixels to ensure that all point sources were
detected.

\section{Analysis}\label{sec:analysis}
\subsection{Point sources}\label{sec:point_sources}

\begin{figure}
\resizebox{\hsize}{!}{\includegraphics{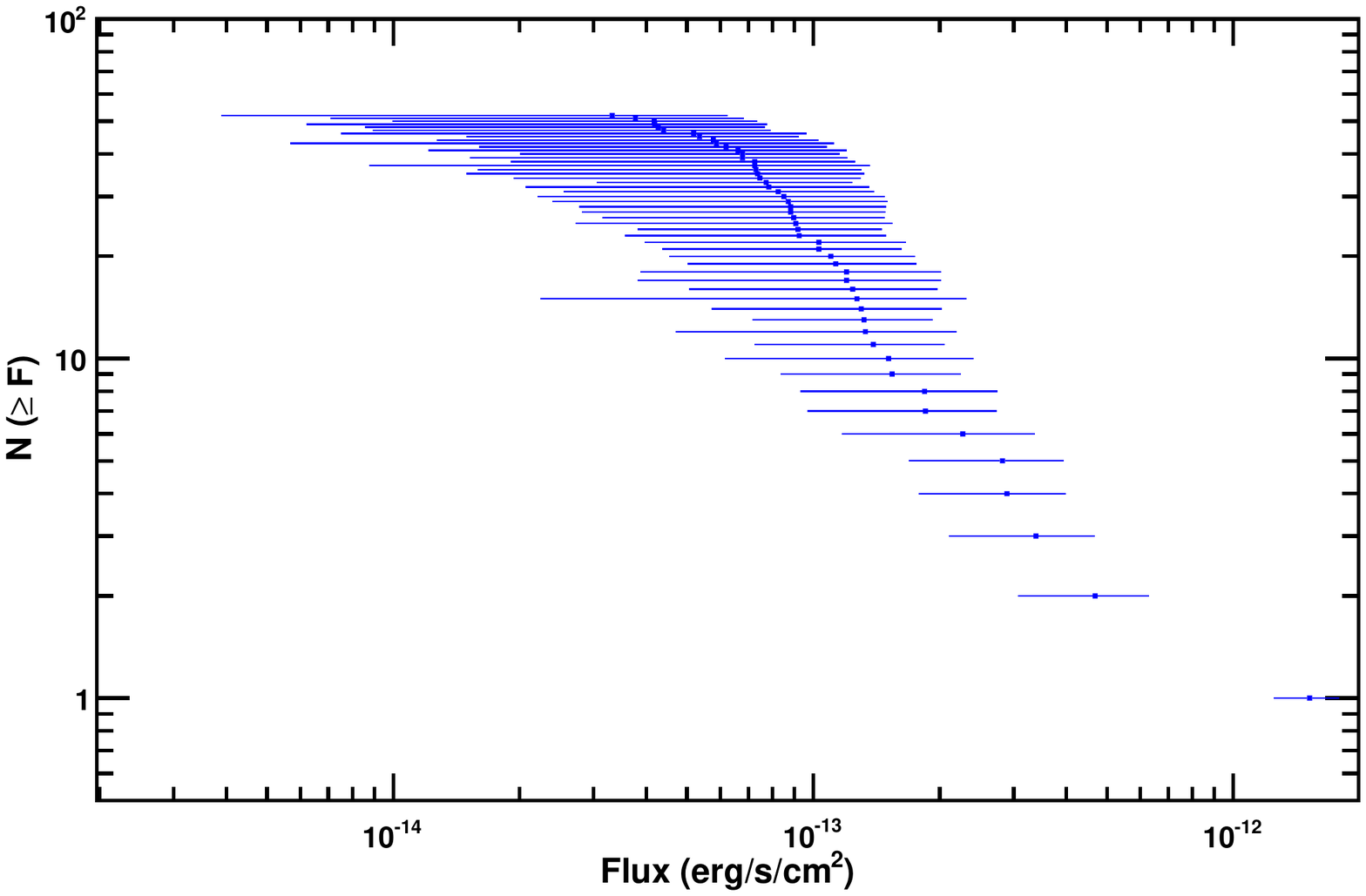}}
\caption{Cumulative $\log{N}-\log{F}$ plot for all five Chandra observations with $N$ being the number of point sources brighter than or equal to flux $F$. All point sources brighter than $7 \times 10^{14}$\,erg\,s$^{-1}$\,cm$^{-2}$ in the $0.5 - 2$\,keV band, assuming a power law with spectral index 2, were removed.}
\label{fig:lognlogf}
\end{figure}

To identify point sources in the field of view (FOV) we analyzed five Chandra snapshot observations matching the five Suzaku pointings. 
The chosen flux limit to remove point sources should be 1.) independent of statistical fluctuations in the source counts and 2.) a compromise between the removed area and accurate treatment of the point sources. Therefore, we made a cumulative $\log N$-$\log F$ plot with $N$ being the number of sources above or equal to a given flux $F$ as shown in Fig. \ref{fig:lognlogf}. When the detection limit of the instrument is reached no more sources are expected to be detected, which results in a flattening of the distribution towards lower fluxes as is the case for $7 \times 10^{-14}$\,erg\,s$^{-1}$\,cm$^{-2}$. All sources brighter than this limit in the $0.5 - 2$\,keV band, assuming a power-law model with spectral index 2, were removed using a circular region of radius $1\arcmin$ around the point source. In Fig. \ref{fig:pointings} all removed point sources are shown.

\begin{figure}
\resizebox{\hsize}{!}{\includegraphics{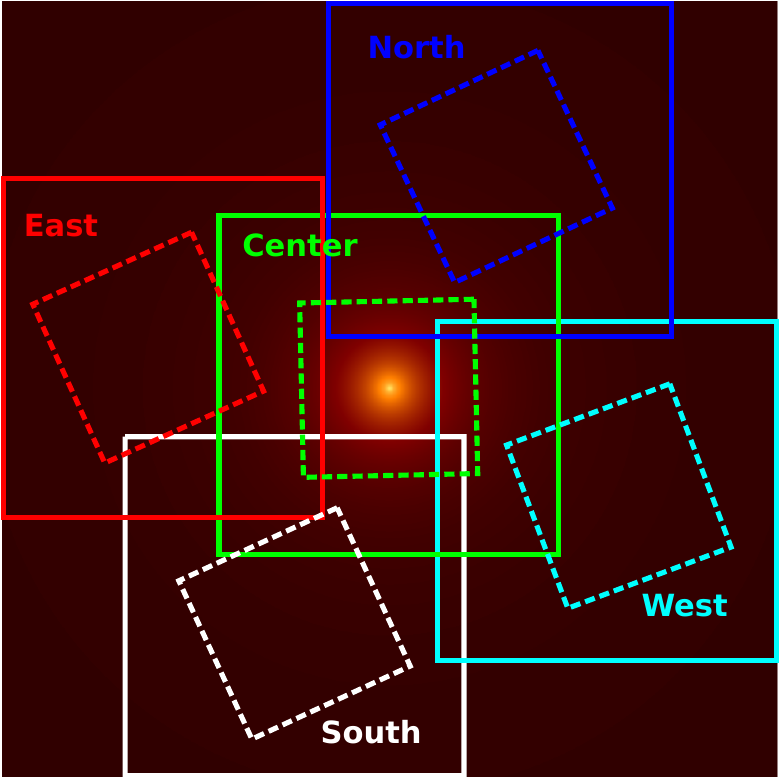}}
\caption{Five Suzaku pointings (dashed) shown together with the underlying double-beta-model surface brightness image obtained with Suzaku and XMM-Newton and the cut-out regions (solid squares) for ARF generation.}
\label{fig:arf_inputs}
\end{figure}

 \subsection{Response files}
 \label{sec:resp_files}
 In a first step, the ancillary response files (ARF) were created using a double-beta model surface brightness image of the galaxy group as was determined using XMM-Newton data (\citealp{2015A&A...573A.118L}) and later iterated using the Suzaku results together with the XMM-Newton data. The profile obtained with XMM-Newton was extrapolated to reach the outskirts of the group. The impact of the input image for the ARF generation on the fit results was investigated and is discussed in Sec. \ref{sec:results}. To ensure sufficient statistics and to save computation time, each observation has its own input image, which is simply a region cut out of the large modeled SB image. This is illustrated in Fig. \ref{fig:arf_inputs} where the five cut-out regions around the central position of each pointing are shown. The regions are larger than the FOV to account for stray light reaching the detector from outside the FOV. The effect of stray light from the galaxy group center can be neglected as is described in Sec. \ref{sec:psf_corr}.
 
 It has to be taken into account that the normalizations and fluxes obtained in the fitting process assume emission from the whole input image (cf. \citealp{2007PASJ...59S.113I}). Therefore, when comparing or linking normalizations of different observations during the fitting process, these parameters have to be rescaled according to the ratio of counts in the corresponding input images used for ARF generation. Also, if it is assumed that the emission spectrum is identical over the whole sky and only the normalization decreases as a function of increasing radius, the XSPEC normalization and flux for each fitted annulus should give the same value as they are rescaled to refer to the whole input image (cf. \citealp{2007PASJ...59S.113I}). However, this assumption does not hold in reality. The flux is a function of temperature and metallicity and, especially for low temperatures, this aspect is not negligible and leads to different normalizations.
 
 For the X-ray background (XRBG), uniform sky ancillary response files, each with a circle of $20\arcmin$ radius, were used. For all ARFs $10^7$ photons were simulated in 157 energy steps using {\it xissimarfgen} and the contamination of the XIS optical blocking filter was taken into account.

 \subsection{PSF correction}\label{sec:psf_corr}
 Suzaku has a PSF of $2\arcmin$ half power diameter which is large enough for photons to be detected in another annulus than the one they truly originate from on the sky. This has influence on the fit results and thus a PSF correction may need to be applied during the fitting process to take the photons actually coming from another annulus into account, as well as stray light. We performed a simulation using the {\it xissim} simulator which is also used for the ARF generation. We generated a photon list of $10^7$ monochromatic X-ray photons of 2\,keV using {\it mkphlist} and determined the photon mixing factors. As input for the simulation we used the same input images as for the ARF generation described earlier. 
 We also performed the simulation for five different energies, but found no significant energy dependence. For the central observation the correction factors are listed in Tab. \ref{tab:psf_factors}. For the outskirts we found that introducing the PSF correction in the fit did not change the fit results significantly -- as is expected owing to the large bin size -- but has influence on the fit stability. Therefore, we neglected the PSF effect in the outskirts. From Tab. \ref{tab:psf_factors} it can be seen that the mixing of photons from the center to the outer annulus is minimal. Therefore we do not expect strong contamination in the outskirts from stray light and PSF effects from the bright galaxy group center.

 \renewcommand{\arraystretch}{1.2}
\begin{table*}
\caption{PSF correction factors for the inner four regions of the central observation (numbered from 1 to 4 outwards) for the final ARF input image (see Sec. \ref{sec:results}). The average of the simulation results for all four detectors is given. The entries should be read as ``Photons coming from annulus 1 on the sky and being detected in annulus 2 on the detector'' indicated as 1 $\longmapsto$ 2 for example.}                   
\centering                                      
\begin{tabular}{cc|cc|cc|cc}          
\hline \hline       
\multicolumn{2}{c|}{To annulus 1} &\multicolumn{2}{c|}{To annulus 2} &\multicolumn{2}{c|}{To annulus 3} &\multicolumn{2}{c|}{To annulus 4} \\\hline
1 $\longmapsto$ 1 & 0.915 & 1 $\longmapsto$ 2 & 0.409 & 1 $\longmapsto$ 3 & 0.110 & 1 $\longmapsto$ 4 & 0.081\\ 
2 $\longmapsto$ 1 & 0.082 & 2 $\longmapsto$ 2 & 0.521 & 2 $\longmapsto$ 3 & 0.225 & 2 $\longmapsto$ 4 & 0.046\\ 
3 $\longmapsto$ 1 & 0.003 & 3 $\longmapsto$ 2 & 0.069 & 3 $\longmapsto$ 3 & 0.618 & 3 $\longmapsto$ 4 & 0.215\\ 
4 $\longmapsto$ 1 & 0.000 & 4 $\longmapsto$ 2 & 0.001 & 4 $\longmapsto$ 3 & 0.047 & 4 $\longmapsto$ 4 & 0.658\\ 

\hline \hline                                             
\end{tabular}
\label{tab:psf_factors}
\end{table*}
\renewcommand{\arraystretch}{1}

 \subsection{Background}\label{sec:background}
 The non-X-ray background (NXB) was determined with {\it xisnxbgen} using Suzaku night Earth data within a time interval of $\pm 150$\,days around the observation date. 
 A careful treatment of the background is important, especially in the outskirts where the group emission is low. The local component of the X-ray background (LHB) is modeled by an unabsorbed apec (astrophysical plasma emission code\footnote{for details about this model see \texttt{www.atomdb.org}}) model with solar metallicity; the temperature is left as a free parameter. The halo component of the background is modeled by an absorbed apec model, also with solar metallicity and a temperature of 0.28\,keV (e.g., \citealp{2010PASJ...62..371H}, \citealp{2011PASJ...63S1019A}). The superposition of extragalactic sources is modeled by an absorbed power law with a spectral index of 1.41 (\citealp{2004A&A...419..837D}). The full XRBG model is then
 ${\rm phabs \times (pow + apec) + apec}.$
 All normalizations are floating in the fit.
 
The estimation of the background parameters is supported by ROSAT all-sky survey data in the energy range from $0.1-2$\,keV. Therefore, the ROSAT spectrum\footnote{obtained with the HEASARC X-ray background tool \texttt{heasarc.gsfc.nasa.gov/cgi-bin/Tools/xraybg/xraybg.pl}}, obtained in an annulus of $0.7-1$\,deg from the center where no group emission is expected, is fitted simultaneously with the Suzaku data, taking into account the different normalizations of the spectra. The background is assumed to be constant across the full analyzed area. 

{As discussed by \citet{2009PASJ...61..805Y}, in some cases a galactic component with higher temperature ($0.4-0.9$\,keV) is needed to describe the X-ray background. We tested for the presence of a higher temperature gas by fitting the ROSAT data with an additional apec component. We found that this model is clearly disfavored by the data comparing the reduced $\chi^2$ values. This is not unexpected as such a model is more often needed at low galactic latitudes, which is not the case for our object. Thus, we performed the analysis with the previously described model without an additional component.}

\subsection{Fitting Strategy}\label{sec:fitting}

For the central observation the quality of the data allowed us to constrain some individual abundances or determine upper limits, especially in the inner annuli. Thus, for the group emission we used a phabs $\times$ vapec\footnote{Because of the PSF correction this model has to be extended to phabs $\times$ (vapec + vapec + vapec + vapec) to account for the contaminating photons that originate from a different annulus on the sky than they are detected in.}
model with the solar abundance table of \citet{2009ARA&A..47..481A}. A vapec model allows individual abundances to be constrained, in contrast to the widely used apec model, which only allows an overall abundance for all elements to be determined. The effect of the chosen abundance table on the fit results is described in Sec. \ref{sec:systematics}. The farther out the annuli lie, the fewer abundance parameters can be constrained by the fit. For this reason, a first fit is performed using an apec model instead of a vapec model in order to determine the average metallicity in each annulus. Then, in a second fit, the abundance parameters of the vapec model, which cannot be constrained by the fit, are fixed to the average value of the first apec fit. The hydrogen column density is fixed to $4.27\times10^{20}$\,cm$^{-2}$ (\citealp{2005yCat.8076....0K}) and the spectral fitting is performed in the energy range 0.5 -- 8.0\,keV.

In the outskirts there is much less group emission than in the center and thus the statistics to constrain parameters is limited. For this reason, the analysis in the outskirts follows two major points:
1.) In a first step one large annulus from $14\arcmin$ -- $34\arcmin$ covering almost the full FOV is analyzed in each of the four observations to check if the group is azimuthally symmetric, and 
2.) If the group turns out to be symmetric, a simultaneous fit including all four observations is performed to reduce the statistical error and the number of annuli can be increased to two.

For the north, east, south, and west observations the group emission is modeled by an absorbed apec model. The temperature, abundance, and normalization of the apec model and also the normalizations of the three background components and the LHB temperature were left as free parameters in the fit. The background temperature for the halo was set to 0.28\,keV (see Sec. \ref{sec:background}). For the simultaneous analyses the same models and parameters were used, but increasing the number of annuli to two in the outskirts (from $14\arcmin$ -- $25\arcmin$ and $25\arcmin$ -- $34\arcmin$) resulting in two apec models, one for each annulus. 

The fitting range was reduced compared to the central observation to 0.8 -- 5.0\,keV because no strong emission from the group is expected at high energies in the outskirts and the impact of the contamination of the XIS optical blocking filter is strongest in the low energy regime. Owing to imperfect calibration between front- and back-illuminated chips, the normalizations for XIS1 were allowed to vary with respect to XIS0 and XIS3 for all models by introducing a multiplicative constant to the model. The value of this constant is on the order of 75\%.

\subsection{Deprojection method}\label{sec:deprojection}

The electron density $n_ {\rm e}$ as a function of radius $R$ is related to the XSPEC normalization $N_i$ in annulus $i$ for an (v)apec model as
\begin{equation}
\label{eq:norm}
N_i = \frac{10^{-14}}{4\pi D_{\rm A}^2(1+z)^2}\int_{V_i} n_{\rm e}(R)n_{\rm H}(R)\,{\rm d}V
\end{equation}
with $n_{\rm H}$ being the hydrogen density and $D_{\rm A}$ the angular diameter distance. 
The emission weighted projected temperature $T^p_i$ {(following \citealp{2002MNRAS.331..635E})} in annulus $i$ is given by
\begin{equation}
\label{eq:temp_deproj}
T^p_i = \frac{\int_{V_i} \epsilon(R)T(R)\,{\rm d}V}{\int_{V_i} \epsilon(R)\,{\rm d}V}
\end{equation}
with emissivity $\epsilon$ and the volume along the line of sight $V_i$. The deprojection of density and temperature was done simultaneously by performing a $\chi^2$ minimization. For the density profile we assumed a single-beta model,
\begin{equation}
 n_{\rm e}(R) = n_0\left(1+\frac{R^2}{R_{\rm c}^2}\right)^{-\frac{3}{2}\beta},
\end{equation} 
where $R_{\rm c}$ is the core radius. 

The temperature is described by a simple power law,
\begin{equation}
 T(R) = AR^b,
\end{equation}
whereas the cool core (the innermost bin) was excluded as it cannot be described by this simplified model. The emissivity is given by $\epsilon = n_{\rm e}n_{\rm H}\Lambda$ with $\Lambda$ being the cooling function which we assume to be constant along the line of sight. More complicated models, which also include the cool core, cannot be used in this case owing to the limited amount of data. However, we find that the temperature profile outside $R>2\arcmin$ is well described by a power law (see Sec. \ref{sec:discussion_temp}) and therefore this model is suited for deprojection.
In the minimization we computed the XSPEC normalization in each annulus using the single-beta model for the electron density, executing the integral in Eq. \ref{eq:norm}, and compared it to the measured normalization in the considered annulus. The same was done simultaneously with the temperature following Eq. \ref{eq:temp_deproj}. The parameters of the single-beta and the power law model were free to vary in the minimization.

We also tested a double-beta model with fixed core radii (taken from the SB fit of the Suzaku and XMM-Newton data), but we found no improvement in the minimization and the $\beta$-values of the two components were approximately the same indicating that a single-beta model is sufficient to reproduce the measured normalizations and temperatures.

The integrated volume $V_i$ corresponding to each annulus $i$ (i.e, the volume along the line of sight) is the cylindrical cut through a sphere with a radius of three times the maximum radius we reach with our observation ($=102\arcmin$). Increasing this radius even further did not change the values of the obtained parameters significantly. The errors were determined using 1000 Monte Carlo (MC) realizations of the measured normalizations and temperatures assuming Gaussian errors and repeating the minimization. For the minimization and the variation in each MC step we take the correlation between all data points into account using the appropriate covariance matrix.

\subsection{Systematics}\label{sec:systematics}
Several sources of systematic uncertainties have been investigated: the chosen abundance table, the uncertainties on the NXB level, and the fluctuation of the CXB due to unresolved point sources. The results of all of these checks are given in Tab.~\ref{tab:results_center}.

Selecting the abundance table of \citet{1989GeCoA..53..197A} instead of \citet{2009ARA&A..47..481A} has a minor influence on the fit results. The values for the iron abundances in the central observation are slightly lower, as are the other abundance values. However, most of them are consistent within the 68\% confidence interval. The lower iron abundances are expected due to the different solar Fe abundances in the two tables. The temperatures are consistent within the uncertainties. 

The NXB background was scaled by $\pm 3\%$  (according to \citealp{2008PASJ...60S..11T}) and the fits were repeated. No strong deviations were observed in the fit results compared to the nominal values {(cf. Tab. \ref{tab:results_center} and \ref{tab:results_outskirts})}.

\citet{2002A&A...389...93L} among others measured a lower value for the MWH gas temperature of 0.2\,keV. Therefore, we tested the influence of fixing this parameter to 0.2\,keV, but found no notable impact on the fit results.

The fluctuations of the CXB due to the statistical fluctuation of the number of point sources in the FOV was measured by \citet{1989PASJ...41..373H} with the Ginga satellite. Following the procedure described in \citet{2013ApJ...766...90I}, the fluctuations were rescaled to the flux limit for point sources used in this analysis and to the analyzed FOV area. 
The fluctuation width is then given by
 
 \begin{equation}\frac{\sigma_{{\rm Suzaku}}}{I_{\rm CXB}} = \frac{\sigma_{\rm Ginga}}{I_{\rm CXB}}\left(\frac{\Omega_{\rm e,Suzaku}}{\Omega_{\rm e,Ginga}}\right)^{-0.5}\left(\frac{S_{\rm c,Suzaku}}{S_{\rm c,Ginga}}\right)^{0.25}
 \label{eq:cxb_fluct}
 \end{equation}
 with $S_{\rm c}$ being the flux limit and $\Omega_{\rm e}$ the effective solid angle of the analyzed region ($\Omega_{\rm e,Ginga} = 1.2$\,deg$^2$). The flux limit for Ginga is $S_{\rm c,Ginga}=6\times 10^{-12}$\,erg\,s$^{-1}$\,cm$^{-2}$ in the $2-10$\,keV band and has been rescaled to the energy band 0.5 -- 2.0\,keV (assuming a power law with spectral index 2.0). The value of $\frac{\sigma_{\rm Ginga}}{I_{\rm CXB}} = 5$ is adopted. The flux limit for our observations determined with Chandra (see Sec. \ref{sec:point_sources}) is $S_{\rm c,Suzaku} = 7 \times 10^{-14}$\,erg\,s$^{-1}$\,cm$^{-2}$ in the 0.5 -- 2.0\,keV band. 
 
 \renewcommand{\arraystretch}{1.2}
\begin{table}
\caption{Fluctuations in the CXB due to statistical fluctuation of the number of point sources in the FOV.}             
\centering                                     
\begin{tabular}{lcccccc}          
\hline \hline                     
Annulus&1&2&3&4&5&6\\
CXB fluctuations (\%) &21.0&11.4&6.4&5.1&2.9&2.6\\
\hline \hline                                         
\end{tabular}
\label{tab:cxb_fluct}
\end{table}
\renewcommand{\arraystretch}{1}

 The resulting values can be found in Tab. \ref{tab:cxb_fluct}. The CXB of each region was scaled according to these values and the fits were repeated. Tables \ref{tab:results_center} and \ref{tab:results_outskirts} show the results. No significant influence on the fit results is observed.\par

\section{Results}\label{sec:results}
In a first step, the impact of the input image used for the ARF generation on the fit results was investigated using an iterative approach. The Suzaku spectral data was fitted using an ARF input image which was created based on the best fit double-beta model for the SB from \citet{2015A&A...573A.118L}. From the results of this spectral fit the SB was recomputed. The next iteration step is to create a new input image based on these fit results. Owing to the limited spatial resolution of Suzaku we additionally use the SB profile from \citet{2015A&A...573A.118L}. From their best fit SB model to XMM-Newton we created pseudo-data that was then fitted simultaneously with the Suzaku results. We use this pseudo-data because the XMM-Newton best fit beta-model profile is corrected for PSF effects (which is not the case for the original data) that is needed in order to fit simultaneously with the PSF corrected Suzaku data. We created pseudo-data up to${~\sim}10\arcmin$ where the signal-to-noise ratio for XMM-Newton approaches one. The simultaneous fit allows us to optimally constrain the surface brightness in the center with XMM-Newton and in the outskirts with Suzaku. With this new SB profile we created a new input image and recreated the ARF files. Using these ARFs we again performed the spectral fits and determined the SB profile with Suzaku. We take into account that the PSF correction changes using the new input image, thus, we repeated the PSF simulation for each new SB profile and used the updated factors in the fit. The profiles after one iteration are shown in Fig. \ref{fig:SB_profile} together with the XMM-Newton pseudo-data. 
The two profiles mostly overlap, especially in the central parts and are in good agreement within the uncertainties. Thus, we conclude that one iteration is sufficient to get a good representation of the actual SB profile of the group in the ARF generation. For further discussion of the SB profile see Sec. \ref{sec:discussion}.

\renewcommand{\arraystretch}{1.2}
\begin{table}
\caption{Fit results for the central observation using an apec emission model. }             
\centering                                      
\begin{tabular}{c c c c}          
\hline \hline                      
Annulus&T [keV]&Z [$Z_{\odot}$]&norm$^*$\\
1&$2.64_{-0.04}^{+0.04}$&$0.94_{-0.06}^{+0.06}$&$2.31_{-0.04}^{+0.04}$\\
2&$3.23_{-0.15}^{+0.15}$&$0.47_{-0.14}^{+0.14}$&$2.17_{-0.04}^{+0.04}$\\
3&$2.45_{-0.17}^{+0.16}$&$0.35_{-0.10}^{+0.12}$&$2.02_{-0.07}^{+0.07}$\\
4&$2.07_{-0.25}^{+0.26}$&$0.41_{-0.18}^{+0.22}$&$1.68_{-0.12}^{+0.14}$\\\hline \hline
\multicolumn{4}{c}{XRBG}\\\hline
&\multicolumn{1}{l}{norm$_{\rm CXB}^\dagger$}&$1.21_{-0.10}^{+0.10}$&\\
&\multicolumn{1}{l}{norm$_{\rm MWH}^{\circ}$}&$0.43_{-0.15}^{+0.15}$&\\
&\multicolumn{1}{l}{$T_{\rm LHB}$($10^{-2}$\,keV)}&$9.87_{-0.47}^{+0.44}$&\\
&\multicolumn{1}{l}{norm$_{\rm LHB}^{\circ}$}&$0.98_{-0.04}^{+0.04}$&\\

\hline \hline                                            
\end{tabular}
\begin{minipage}{\columnwidth}
\vspace{0.2cm}
\footnotesize $^*$ ${\rm norm}=\frac{1}{4\pi[D_A(1+z)]^2}\int n_{\rm e}n_{\rm H}{\rm d}V\,10^{-16}\,{\rm cm}^{-5}$ with $D_A$ being the angular diameter distance to the source.\\
$^{\circ}$ Normalization of the apec component scaled to area 400$\pi$ assumed in the uniform-sky ARF calculation. \\ \phantom{-,}norm = $\frac{1}{4\pi[D_A(1+z)]^2}\int n_{\rm e}n_{\rm H}{\rm d}V$\,10$^{-20}$\,cm$^{-5}$ .
\end{minipage}
\begin{minipage}{\columnwidth}
\vspace{0.03cm}
\footnotesize  $^\dagger$ in units of $10^{-3}$ photons/s/cm$^2$ at 1\,keV scaled to the area 400$\pi$.
 \end{minipage}
\label{tab:results_apec}
\end{table}
\renewcommand{\arraystretch}{1}

\begin{figure}
\resizebox{\hsize}{!}{\includegraphics{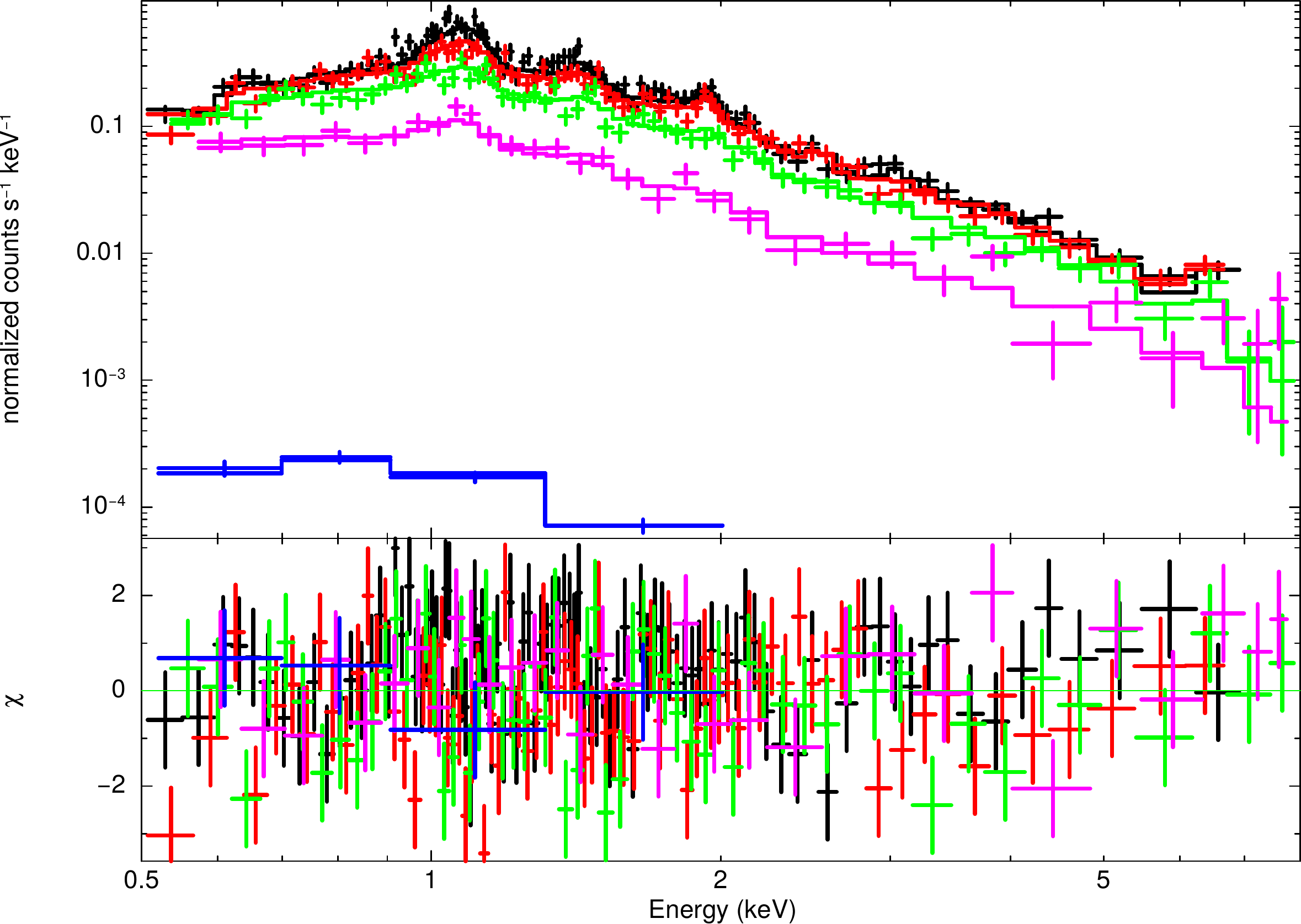}}
\caption{Suzaku spectra for XIS1 and best fit models (solid lines) and residuals of all four regions in the central observation. The lowermost blue data points correspond to the ROSAT spectrum of the XRBG, which also extends to lower energies (not shown).}
\label{fig:vapec_spectra}
\end{figure}

As described in Sec. \ref{sec:fitting}, the first fit to the data of the central observation was performed using an apec model. The results for this fit are given in Tab. \ref{tab:results_apec}. The abundance values determined in this fit were then used to fix the indeterminable abundance parameters of the vapec model in the second fit. The results for the second fit are given in Tab. \ref{tab:results_center}. 
We measured individual abundances for Mg, Si, S, Ar, and Ca whereas some of these can only be constrained in the inner annuli or only upper limits are given (cf. Tab. \ref{tab:results_center}). The temperature for the local background component agrees well with \citet{2010PASJ...62..371H} and \citet{2011PASJ...63S1019A}. The CXB intensity in the $2 - 10$\,keV band is $2.30_{-0.17}^{+0.18}\times 10^{-11} {\rm erg}\,{\rm s}^{-1}\,{\rm cm}^{-2}\,{\rm deg}^{-2}$ and in good agreement with measurements by, e.g., \cite{2002A&A...389...93L} and \cite{2004A&A...419..837D}. The spectra together with the best fit vapec models are shown in Fig. \ref{fig:vapec_spectra}. No strong residuals can be seen and the reduced $\chi^2$ is $1.2$.

\renewcommand{\arraystretch}{1.2}
\begin{table}
\caption{Fit results for the north, east, south, and west observations in the region $14\arcmin$ -- $34\arcmin$. }              
\centering                                      
\begin{tabular}{l | c | c | c | c}          
\hline \hline                     
& north & east & south & west \\ \hline 

$T$ (keV) &$ 1.14 _{ -0.10 }^{+ 0.12 }$ & $ 1.08 _{ -0.05 }^{+ 0.11 }$ & $ 1.16 _{ -0.16 }^{+ 0.12 }$ & $ 1.33 _{ -0.08 }^{+ 0.20 }$ \\
$Z$ ($Z_\odot$) &$ 0.38 _{ -0.61 }^{+ 0.80 }$ & $ 0.37 _{ -0.52 }^{+ 0.06 }$ & $ 0.29 _{ -0.44 }^{+ 0.13 }$ & $ 0.23 _{ -0.31 }^{+ -0.11 }$ \\
norm$^*$&$ 1.45 _{ -1.01 }^{+ 1.08 }$ & $ 1.91 _{ -0.91 }^{+ 0.89 }$ & $ 1.90 _{ -0.95 }^{+ 1.00 }$ & $ 3.60 _{ -0.82 }^{+ 0.82 }$ \\ \hline  \hline
 \multicolumn{5}{c}{XRBG}\\\hline
norm$_{\rm CXB}^\dagger$ &$ 1.26 _{ -0.05 }^{+ 0.05 }$ & $ 1.25 _{ -0.05 }^{+ 0.05 }$ & $ 1.28 _{ -0.05 }^{+ 0.05 }$ & $ 1.14 _{ -0.06 }^{+ 0.05 }$ \\
norm$_{\rm MWH}^{\circ}$ &$ 4.41 _{ -1.23 }^{+ 1.24 }$ & $ 4.62 _{ -1.33 }^{+ 1.31 }$ & $ 5.17 _{ -1.54 }^{+ 1.21 }$ & $ 7.03 _{ -1.22 }^{+ 1.34 }$ \\
$T_{\rm LHB}$($10^{-2}$\,keV) &$ 9.87 _{ -0.47 }^{+ 0.44 }$ & $ 9.84 _{ -0.47 }^{+ 0.45 }$ & $ 9.74 _{ -0.48 }^{+ 0.45 }$ & $ 9.78 _{ -0.47 }^{+ 0.45 }$ \\
norm$_{\rm LHB}^{\circ}$ &$ 9.81 _{ -0.43 }^{+ 0.43 }$ & $ 9.76 _{ -0.43 }^{+ 0.43 }$ & $ 9.63 _{ -0.43 }^{+ 0.43 }$ & $ 9.76 _{ -0.43 }^{+ 0.43 }$ \\

\hline \hline                                            
\end{tabular}
\begin{minipage}{\columnwidth}
\vspace{0.2cm}
\footnotesize $^*$ ${\rm norm}=\frac{1}{4\pi[D_A(1+z)]^2}\int n_{\rm e}n_{\rm H}{\rm d}V\,10^{-16}\,{\rm cm}^{-5}$ with $D_A$ being the angular diameter distance to the source and rescaled to the central observation for better comparability.\\
$^{\circ}$ Normalization of the apec component scaled to area 400$\pi$ assumed in the uniform-sky ARF calculation. \\ \phantom{-,}norm = $\frac{1}{4\pi[D_A(1+z)]^2}\int n_{\rm e}n_{\rm H}{\rm d}V$\,10$^{-20}$\,cm$^{-5}$ .
\end{minipage}
\begin{minipage}{\columnwidth}
\vspace{0.03cm}
\footnotesize  $^\dagger$ in units of $10^{-3}$ photons/s/cm$^2$ at 1\,keV scaled to the area 400$\pi$.
 \end{minipage}
\label{tab:results_nesw}
\end{table}
\renewcommand{\arraystretch}{1}

\begin{figure}
\resizebox{\hsize}{!}{\includegraphics{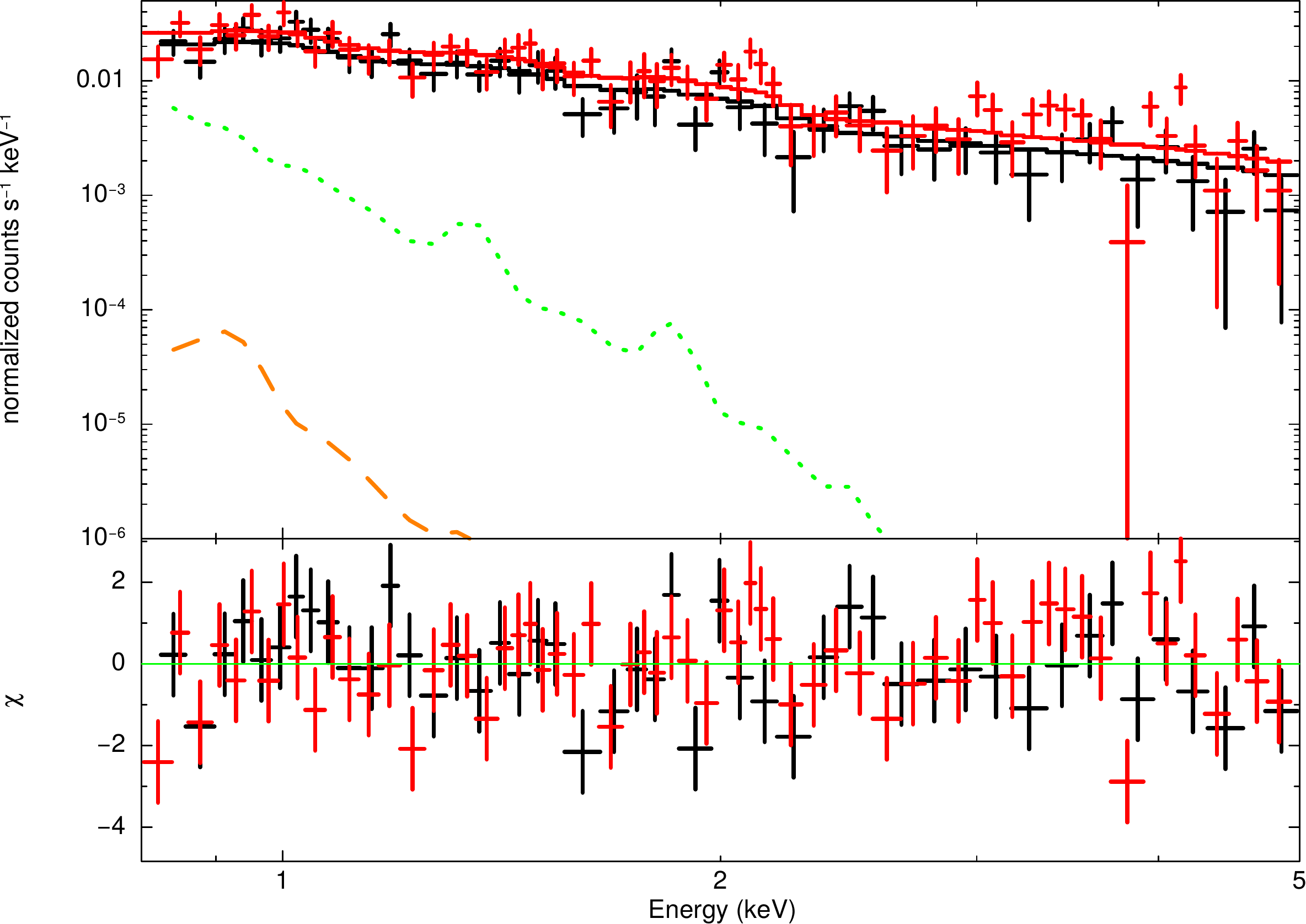}}
\caption{Suzaku spectra and residuals of the two outskirts regions for the north observation and XIS1. The solid lines show the best fit model, the orange dashed line shows the model component for the local background, and the green dotted line corresponds to the halo background model component.}
\label{fig:spectra_aussen}
\end{figure}

The fit results for individual fits of the north, east, south, and west observations are given in Tab. \ref{tab:results_nesw}. For better comparability the normalizations have been rescaled to match the central input image which is necessary owing to the different input images during ARF generation. 
The values for temperatures, abundances, and normalizations are mostly consistent within one standard deviation. Only the western observation shows slightly higher temperature and normalization but this deviation is not significant (less than 2$\sigma$). Of course, the uncertainties are quite large due to the limited statistics in the outskirts which we improved in a simultaneous fit of all outer observations. The values for the background parameters agree within the uncertainties and gives us confidence in our treatment of the background. For these reasons we conclude that the galaxy group is sufficiently symmetric in the azimuthal directions so that a simultaneous fit of all outskirts observations is justified. The increased statistics in the fit allows for splitting the outer region from $14\arcmin$ -- $34\arcmin$ into two annuli from $14\arcmin$ -- $25\arcmin$ and $25\arcmin$ -- $34\arcmin$. 
{The results for this fit are given in Tab. \ref{tab:results_outskirts_mainpart}}. 
The corresponding spectra are shown in Fig. \ref{fig:spectra_aussen} where no strong residuals are visible.

\renewcommand{\arraystretch}{1.2}
\begin{table}
\caption{Results for the simultaneous fit to the outskirts observations. }              
\centering                                      
\begin{tabular}{c c  c  c}          
\hline \hline                     
Annulus&T [keV]&Z [$Z_{\odot}$]&norm$^*$\\
5&$1.20_{-0.10}^{+0.07}$&$0.28_{-0.11}^{+0.17}$&$1.48_{-0.39}^{+0.43}$\\
6&$1.18_{-0.09}^{+0.07}$&$0.39_{-0.15}^{+0.29}$&$1.84_{-0.64}^{+0.67}$\\
\hline \hline
\multicolumn{4}{c}{XRBG}\\\hline
&\multicolumn{1}{l}{norm$_{\rm CXB}^\dagger$}&$1.24_{-0.03}^{+0.03}$&\\
&\multicolumn{1}{l}{norm$_{\rm MWH}^{\circ}$}&$6.32_{-0.96}^{+1.91}$&\\
&\multicolumn{1}{l}{$T_{\rm LHB}$($10^{-2}$\,keV)}&$9.71_{-0.48}^{+0.93}$&\\
&\multicolumn{1}{l}{norm$_{\rm LHB}^{\circ}$}&$9.65_{-0.43}^{+0.85}$&\\

\hline \hline                                            
\end{tabular}
\begin{minipage}{\columnwidth}
\vspace{0.2cm}
\footnotesize $^*$ ${\rm norm}=\frac{1}{4\pi[D_A(1+z)]^2}\int n_{\rm e}n_{\rm H}{\rm d}V\,10^{-16}\,{\rm cm}^{-5}$ with $D_A$ being the angular diameter distance to the source and rescaled to the central observation for better comparability.\\
$^{\circ}$ Normalization of the apec component scaled to area 400$\pi$ assumed in the uniform-sky ARF calculation. \\ \phantom{-,}norm = $\frac{1}{4\pi[D_A(1+z)]^2}\int n_{\rm e}n_{\rm H}{\rm d}V$\,10$^{-20}$\,cm$^{-5}$ .
\end{minipage}
\begin{minipage}{\columnwidth}
\vspace{0.03cm}
\footnotesize  $^\dagger$ in units of $10^{-3}$ photons/s/cm$^2$ at 1\,keV scaled to the area 400$\pi$.
 \end{minipage}
\label{tab:results_outskirts_mainpart}
\end{table}
\renewcommand{\arraystretch}{1}

As expected from the individual fits, the temperature in the outskirts is slightly higher than ${\sim}1$\,keV. {The abundance in both regions is consistent within the uncertainties}. The background parameters are all consistent with the ones from the individual analyses and the reduced $\chi^2$ is 1.1 which gives us confidence that azimuthal symmetry is a good assumption.

\section{Discussion}\label{sec:discussion}
\subsection{Surface brightness profile}\label{sec:discussion_sb}
\begin{figure}
\resizebox{\hsize}{!}{\includegraphics{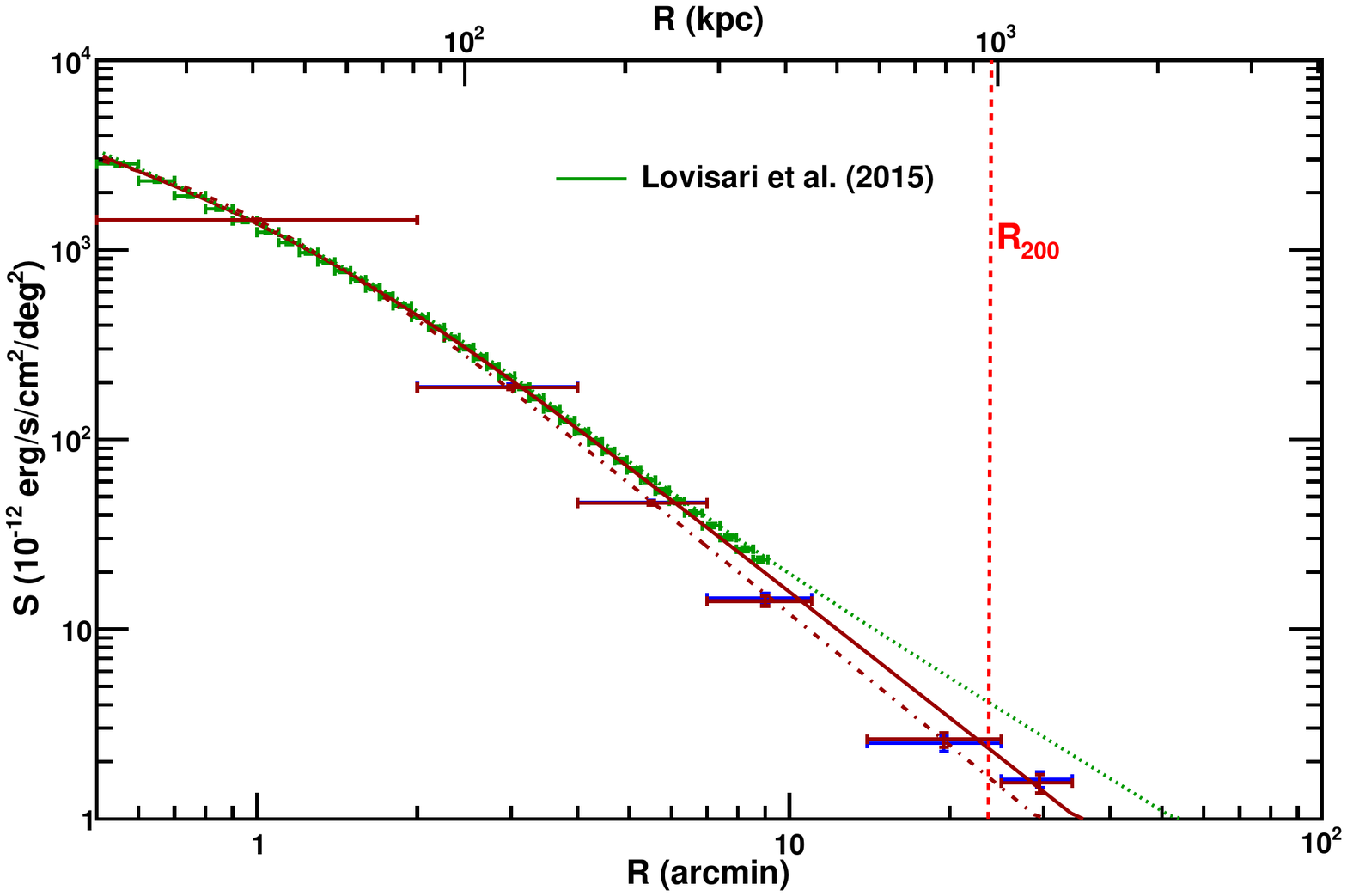}}
\caption{Surface brightness profile of UGC03957 in the 0.7 -- 2\,keV band for different input images used for the ARF creation. Green data points correspond to XMM-Newton pseudo-data and the green dotted line shows the extrapolated SB profile from \citet{2015A&A...573A.118L}. Blue data points correspond to the measured SB when an input image following the extrapolated XMM-Newton results is used. Red data points correspond to the SB using an ARF input image created from the best fit to the measured Suzaku SB from the first iteration step and the XMM-Newton pseudo-data simultaneously. The red dot-dashed line shows the best fit single-beta model to the red data points; the red solid line represents the best fit double-beta model to the red data points and XMM-Newton pseudo-data simultaneously.}
\label{fig:SB_profile}
\end{figure}

\citet{1999ApJ...525...47V} found that a single-beta model profile with $\beta = 0.65-0.85$ accurately describes the surface brightness profiles for 39 local clusters in the range $(0.3 - 1)\,R_{180}$ measured with ROSAT PSPC. However, e.g. \citet{1999A&A...344..755K} and \citet{1999ApJ...516..604H} measured flatter profiles for galaxy groups with $\beta \sim 0.4-0.5$. When we perform a single-beta model fit to the Suzaku SB data (dot-dashed line and red data points in Fig. \ref{fig:SB_profile}) we obtain $\beta=0.55\pm0.01$, which is in agreement with the results for galaxy groups. \citet{2007MNRAS.377..595K} among others measured the SB profiles of a sample of fossil groups and found values of $\beta=0.43-0.60$, which is similar to our findings. These smaller $\beta$-values compared to galaxy clusters have already been seen in early ROSAT results from, e.g., \citet{1995ApJ...445..578D} and \citet{1995ApJ...449..422H}. The former studied three groups as part of a larger cluster sample and found $\beta=0.38-0.53$, whereas the clusters gave higher values between $\beta=0.53-0.74$. \citet{1995ApJ...449..422H} found comparable results for their study of four galaxy groups. \citet{1995AJ....110...46D} measured the SB profile for five poor clusters with ROSAT PSPC and also found values between $\beta=0.47-0.60$. These results all clearly show that there is a deviating behavior of galaxy groups compared to galaxy clusters concerning the surface brightness. For our measurement (cf. Fig. \ref{fig:SB_profile}) we note that the last data point shows a weak indication of a flattening of the profile towards larger radii. This flattening could be reflected in the density profile (cf. Sec. \ref{sec:discussion_dens}) and has been observed previously with Suzaku by, e.g., \citet{2010ApJ...714..423K}. They found a flattening in the density profile for a Suzaku observation of the cluster A1689. A flatter profile affects the hydrostatic mass estimate and can result in lower total cluster masses. {We note that a flattening has also been observed by other Suzaku studies. \citet{2013ApJ...775...89S} measured a clearly higher density in the outskirts compared to their best fit single-beta model for a fossil group. \citet{2011Sci...331.1576S} also measured a flattening of the density profile for the north-east direction of the Perseus cluster. One explanation for the flattening is gas clumping in the outskirts of clusters and groups which can lead to an overestimate in the gas density. \citet{2011ApJ...731L..10N} performed hydrodynamical simulations of 16 galaxy clusters and studied the effect of gas clumping in the outer parts. Their results suggest that this effect is not negligible when dealing with cluster outskirts. Therefore, it is very important to have accurate calibration of the instruments and more studies reaching large radii to deduce whether the flattening is an instrumental effect or a real property of the gas in the outskirts of groups and clusters.}

However, for our measured SB profile, the simultaneous double-beta model fit to XMM-Newton pseudo-data and Suzaku prefers a slightly flatter profile in the outskirts, indicating that XMM-Newton tends to larger SB values compared to Suzaku when going to larger radii. This is also represented by the green dashed line, which shows the extrapolated XMM-Newton profile obtained by \citet{2015A&A...573A.118L}. The extrapolated profile clearly differs from the Suzaku measurement in the outer parts of the group and emphasizes the importance of having accurate measurements out to large radii to avoid biases in the calculations due to extrapolation.

\begin{figure}
\resizebox{\hsize}{!}{\includegraphics{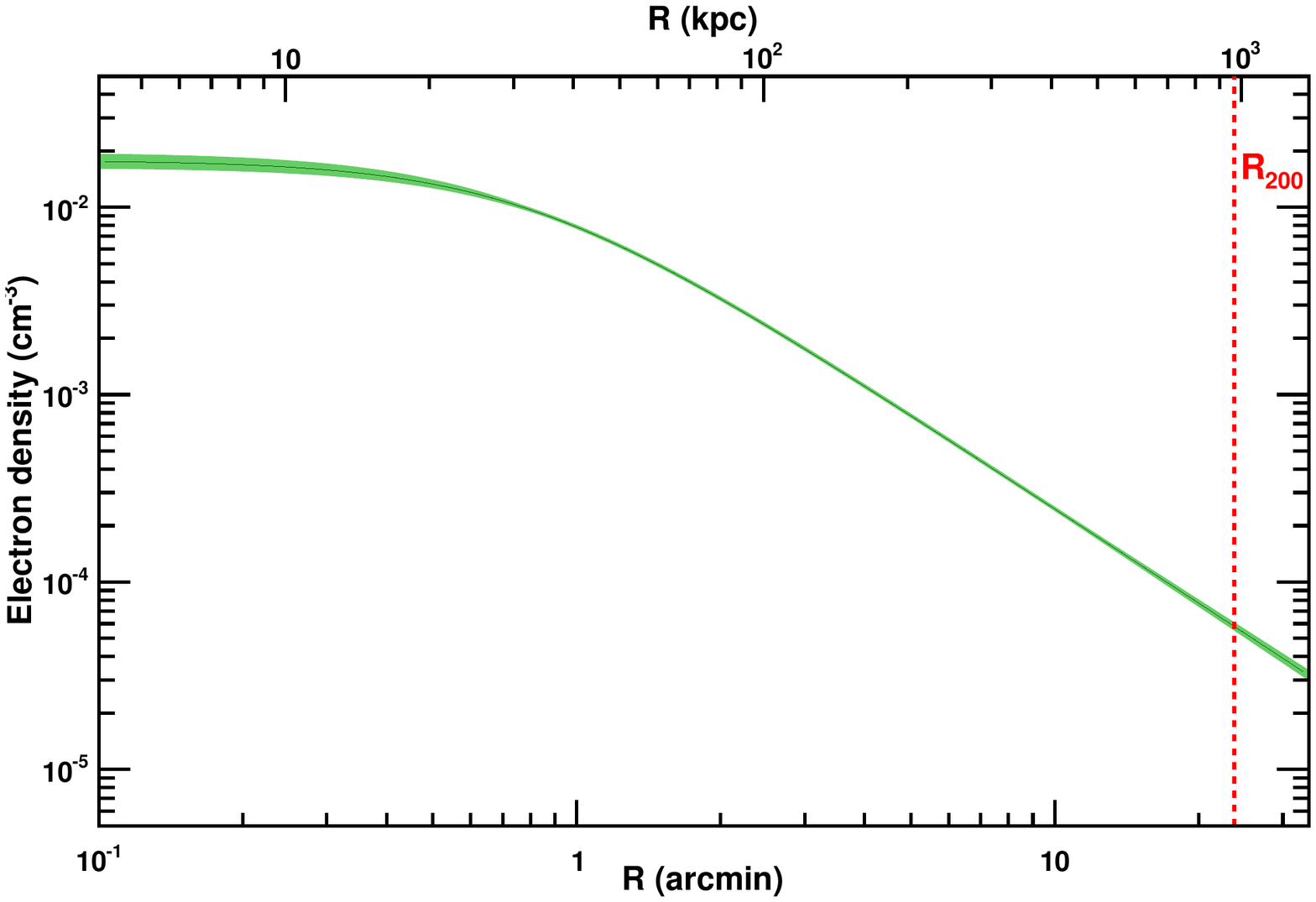}}
\caption{Deprojected density profile of UGC03957. Shaded area corresponds to the formal 90\% confidence region. For a description of the deprojection method see Sec. \ref{sec:deprojection}.}
\label{fig:dens_profile}
\end{figure}

\begin{figure}
\resizebox{\hsize}{!}{\includegraphics{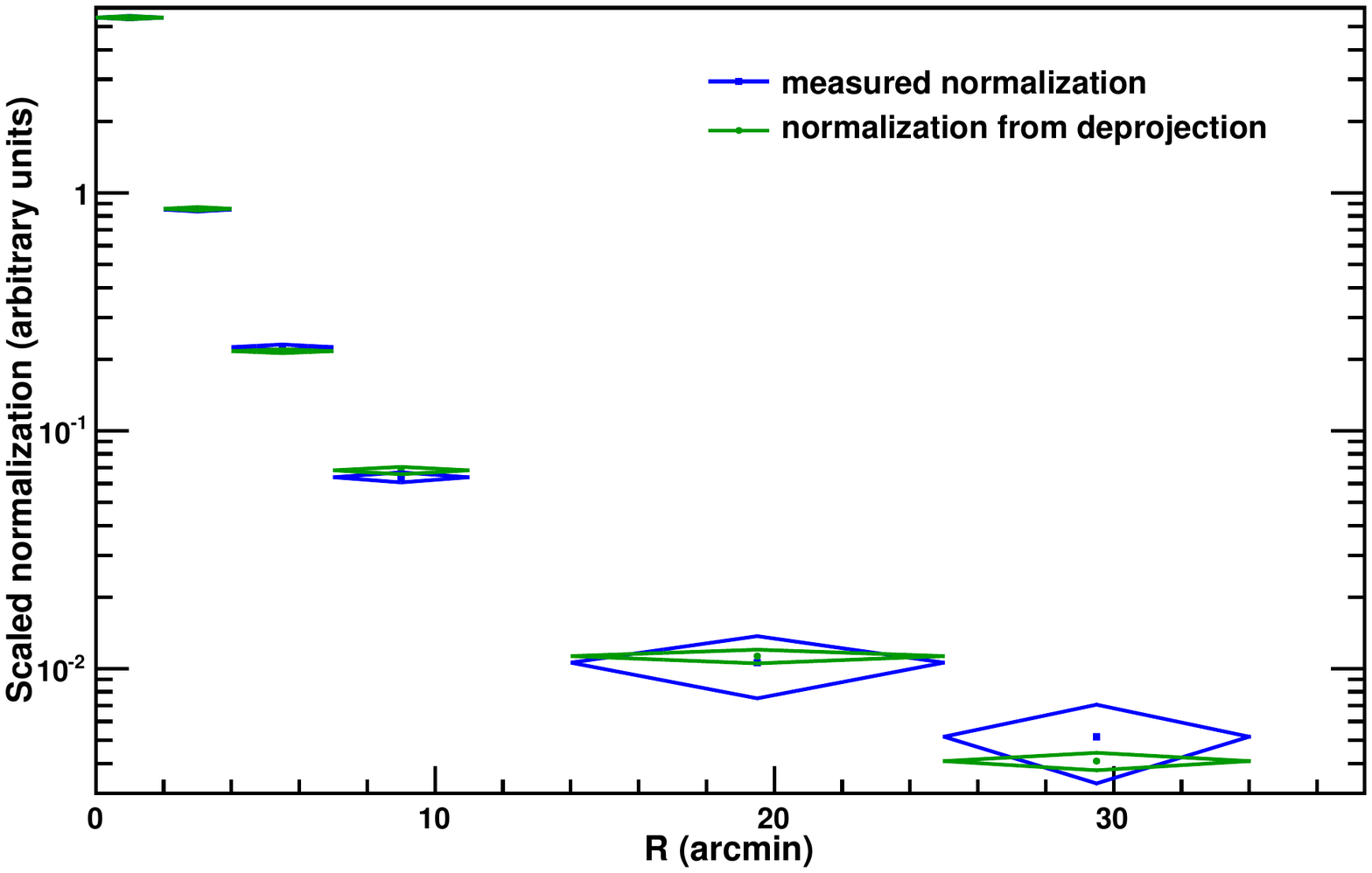}}
\caption{Normalizations of the vapec model scaled by the integrated volume $V_i$ (c.f. Eq. \ref{eq:norm}) along the line of sight in arbitrary units. Blue diamonds correspond to the measured normalizations, while green diamonds show the normalizations obtained from the deprojected density profile.}
\label{fig:norms}
\end{figure}

\subsection{Density profile}\label{sec:discussion_dens}
The deprojected density profile is shown in Fig. \ref{fig:dens_profile}. The best fit parameters for the single-beta model can be found in Tab. \ref{tab:density_params}. 
For the deprojection we reconstructed the XSPEC normalizations using Eq. \ref{eq:norm}. The statistical uncertainties are small owing to the small statistical errors of the XSPEC normalization (see Tab.~\ref{tab:results_center}) which dominate the shape of the beta model. {We note that especially for a low temperature plasma the normalization and abundance parameter of the apec model are correlated. Therefore, we checked whether fixing the abundance leads to significantly different normalizations. We tested for two cases: 1.) fixing the abundance to 0.3\,$Z_\odot$ for both regions and 2.) a more extreme case, fixing the parameters to 0.4\,$Z_\odot$ and 0.2\,$Z_\odot$ for regions 5 and 6, respectively. The first case yields similar results to the nominal fit for temperature and normalization. For the second case we find a ${\sim}$35\,\% higher normalization in the outermost region however, all the values are consistent within the uncertainties with the results when the abundance parameters are left as free parameters in the fit. Thus, we proceed using the results from the latter fit as in this case the statistical uncertainties on the abundance can be taken into account for the further analysis.}

The measured normalizations from each annulus and the normalizations determined from the minimization are shown in Fig. \ref{fig:norms} normalized to the integrated volume $V_i$ (cf. Eq. \ref{eq:norm}) in arbitrary units. The comparison shows that our measurement can be reproduced well by a single-beta model profile. {The last data point is slightly lower than the measured XSPEC normalization however, the indication of  a flattening is weak (see also the SB profile in Fig. \ref{fig:SB_profile})}. If the determined deprojected density profile in the outskirts is slightly steeper than the actual profile, the normalizations determined in the minimization will lead to a lower value than the observed one. The indication for a flattening is not significant and in our case, the overall profile is reproduced well by our method using a single-beta model.  As mentioned above, a systematic flattening would have significant impact on the mass estimates. As it is difficult to get robust constraints in the outskirts and many analyses are limited to $R_{500}$, density profiles are often extrapolated to larger radii. If the actual density profile is flatter in the outer parts, this extrapolation results in an overestimation of the total mass or, on the other hand, if non-gravitational effects such as clumping bias gas density measurements in the outskirts towards higher values this would cause an underestimation of the cluster mass. These effects have a direct influence on the determination of cosmological parameters and could cause biases. 
Additionally, we need to know which other non-gravitational effects might affect the measurements. In addition to clumping, non-equilibrium states such as deviations from thermal equilibrium between protons and electrons might also be present as suggested by measurements from, e.g., \citet{2011PASJ...63S1019A}. See \cite{2013SSRv..tmp...33R} for a review of these effects. This can be best tested using the entropy profile and the gas mass fraction, which we investigate in Sec. \ref{sec:discussion_mass}.

\renewcommand{\arraystretch}{1.2}
\begin{table}
\caption{Single-beta model parameters for the deprojected density profile.}           
\centering                                      
\begin{tabular}{c c}          
\hline \hline                      
$n_0$ (cm$^{-3}$) & $(1.77^{-0.09}_{+0.10})\times 10^{-2}$\\
$R_{\rm c}$ (arcmin) &$0.78_{-0.05}^{+0.04}$\\
$\beta$ & $0.56^{+0.01}_{-0.01}$\\
\hline \hline                                             
\end{tabular}
\label{tab:density_params}
\end{table}
\renewcommand{\arraystretch}{1}

\subsection{Temperature profile}\label{sec:discussion_temp}
\begin{figure}
\resizebox{\hsize}{!}{\includegraphics{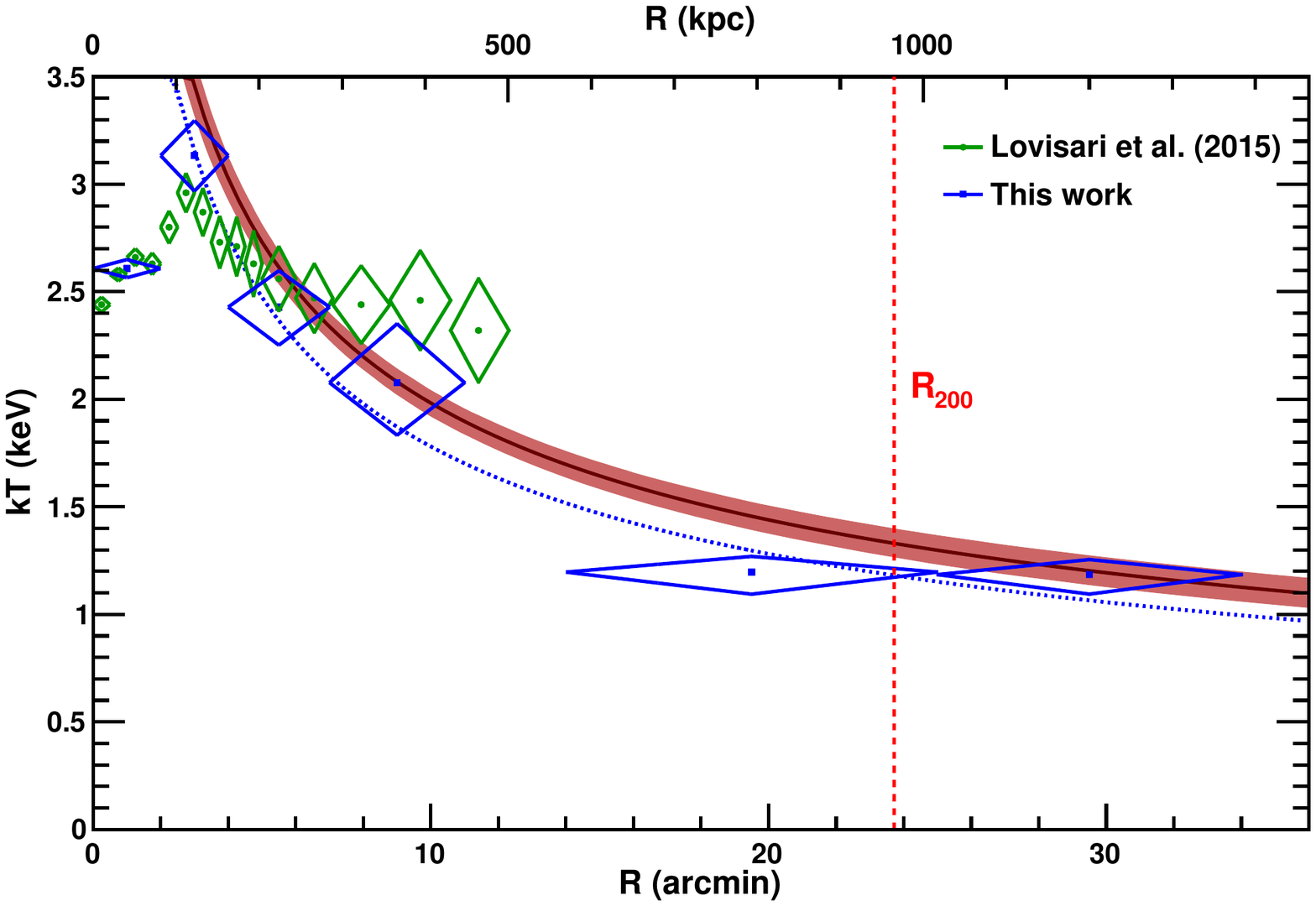}}
\caption{Temperature profile of UGC03957 measured with Suzaku (blue) and XMM-Newton (green). The blue dotted line shows the best fit power law to the Suzaku data excluding the cool core. The red solid line corresponds to the deprojected temperature profile, and the shaded area shows the 68\% uncertainties. Uncertainties on the measurements in the inner four annuli are larger due to the shorter exposure time (cf. Tab. \ref{tab:observations}).}
\label{fig:temp_profile}
\end{figure}

The temperature profiles for UGC03957 measured with Suzaku and by \citet{2015A&A...573A.118L} with XMM-Newton are shown in Fig. \ref{fig:temp_profile}. Both profiles clearly show the cool core of the group and are in very good agreement within the uncertainties. Nevertheless, we note that XMM-Newton tends to higher values around $R{\sim} 10\arcmin$ which would lead to a bias if the profile is extrapolated to larger radii. 

A temperature drop of a factor of ${\sim} 3$ from the center to the outskirts of the group is consistent with previous Suzaku measurements of galaxy clusters (cf. Fig. 9 of \citealp{2013SSRv..tmp...33R}). The solid lines in Fig. \ref{fig:temp_profile} correspond to the best fit power-law model to the projected Suzaku data points (blue line) and the deprojected profile (red line). The innermost bin was excluded. As expected from the negative temperature gradient, we see that the deprojected temperature profile is slightly higher than the projected one. In the following the deprojected temperature profile is used to compute mass and entropy.

\subsection{Abundance and supernova ratio}\label{sec:discussion_abund}
\begin{figure}
\resizebox{\hsize}{!}{\includegraphics{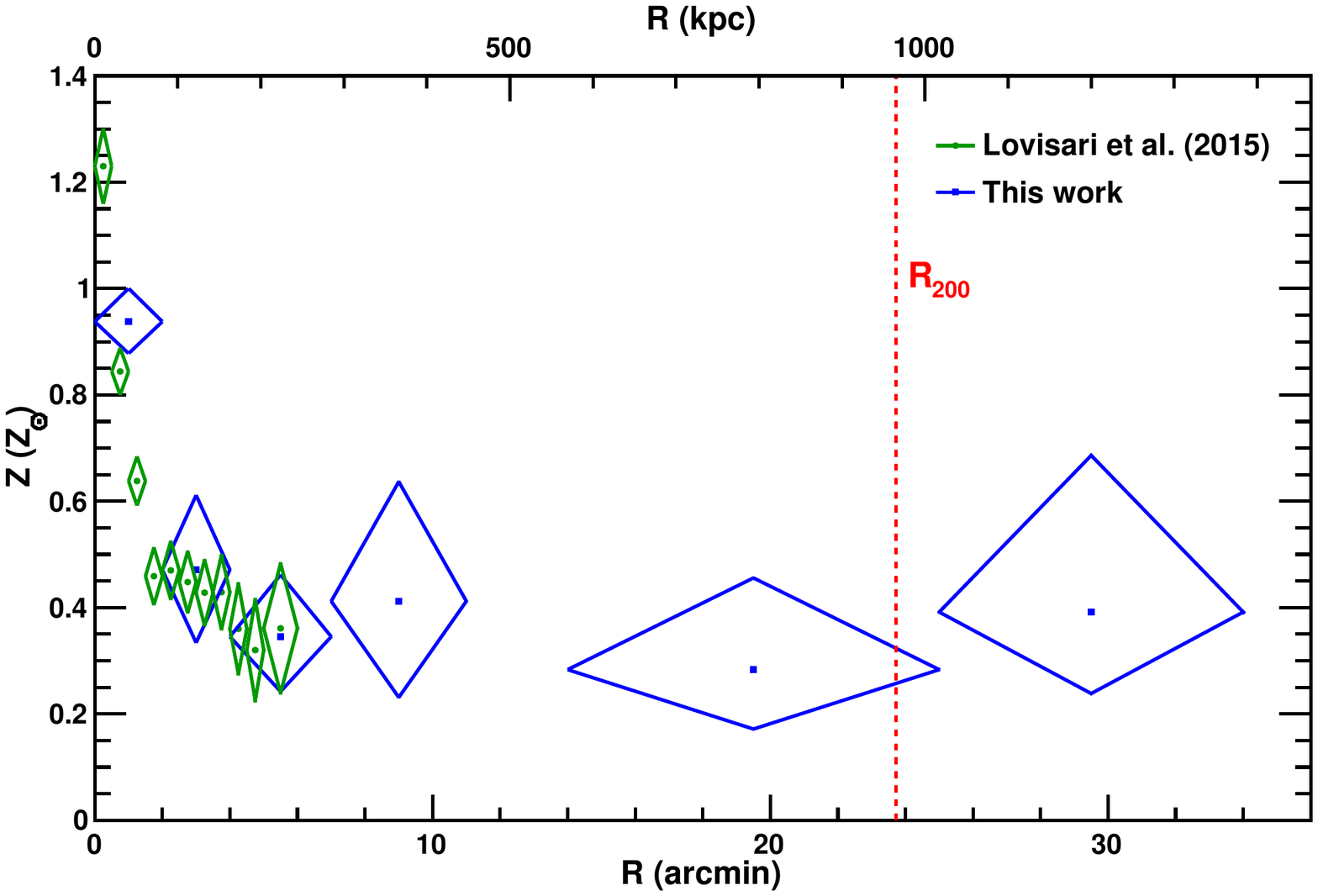}}
\caption{Abundance profile of UGC03957 measured with Suzaku (for values see Tab. \ref{tab:results_apec}) in blue and XMM-Newton (\citealp{2015A&A...573A.118L}) in green.}
\label{fig:abund_profile}
\end{figure}

Figure \ref{fig:abund_profile} shows the abundance profile using the average abundances determined in the apec fit for the central observation and the abundance measured in the outer observations reaching beyond $R_{200}$. The abundance drops from the innermost to the second bin and then shows a rather flat behavior out to the outskirts. The profile is in good agreement with the XMM-Newton measurements where they overlap in the inner parts. 

\citet{2007A&A...466..813K} simulated two possible mechanisms for the enrichment of the ICM: ram pressure stripping and galactic winds. Ram pressure stripping is most effective at high densities, thus in the center of galaxy groups and clusters, whereas it is expected to have less influence at the outer parts. Galactic winds are more effective in lower density regions because of the lower pressure of the surrounding material. In their simulation \citet{2007A&A...466..813K} showed that when ram pressure stripping is the primary process a steeper abundance profile is expected than for galactic winds. Therefore, the flat profile in our measurement is a hint that galactic winds are the dominant enrichment process outside the group center. This is consistent with the first abundance measurements out to the virial radius of two galaxy clusters (\citealp{2008PASJ...60S.343F}). From the central to the second bin a steep gradient is observed. Here the impact of the brightest central galaxy, which probably contributes significantly to the enrichment, is an important factor.

We also studied individual abundances in the inner annuli of the central observation. We note a high Ar and Ca abundance in the innermost annulus (albeit with high uncertainties), yielding Ar/Fe $=2.6\pm0.7$ and Ca/Fe $=1.3^{+0.6}_{-0.5}$, respectively.
{\cite{2009A&A...493..409S} found comparable high values for Ca/Fe in the central region of the Hydra A cluster. Also \citet{2007A&A...465..345D} measured individual abundances for a sample of 22 galaxy clusters observed with XMM-Newton and found high Ca/Fe values of ${\sim} 1.5$\,Z$_{\odot}$ for several clusters. However, they found lower Ar/Fe values in the central parts, whereas a stacked analyses of all archival X-ray ASCA data performed by \citet{2005ApJ...620..680B} yield comparable high Ar abundance for the low temperature systems. Our measured abundance values for Mg, Si, and S in the central parts are in good agreement with Suzaku measurements by, e.g., \citet{2007PASJ...59..299S} and \citet{2008PASJ...60S.317T}, and by \citet{2009PASJ...61S.337K} who also studied a galaxy group.}

The ratio of SNIa and SNCC that exploded in the past can be determined using the abundances of $\alpha$-elements such as Si and S compared to Fe. We measured the abundance of these elements at intermediate radii between $2\arcmin-11\arcmin$ to minimize a possible influence of the central galaxy (indicated by the steep gradient in Fig. \ref{fig:abund_profile}) yielding  ${Z_{\rm Fe} = 0.39_{-0.06}^{+0.06}\,Z_\odot}$, $Z_{\rm Si} = 0.46_{-0.15}^{+0.16}\,Z_\odot$, and $Z_{\rm S} = 0.70_{-0.24}^{+0.25}Z_\odot$. Then we followed the procedure described by \citet{2009A&A...508..191L} {and determined the SN ratio for each of the two elements}. Two models for the yields of SNIa were tested: a deflagration model (W7-model) and a delayed detonation model (WDD2), as described in \citet{1999ApJS..125..439I}. Average SNCC yields in the mass range of $10{M_\odot}$ to $50{M_\odot}$, calculated by \citet{1995MNRAS.277..945T} assuming a Salpeter initial mass function, were used. {The SN ratio is defined as $R=N_{\rm SNCC}/(N_{\rm SNCC}+N_{\rm SNIa})$, where $N$ is the number of SN for a given type.}

The results are given in Tab. \ref{tab:sn_ratio}. Both models yield similar results and are consistent within the uncertainties. The SN ratios for Si and S also match within the uncertainties; thus, the observed abundances for UGC03957 can be explained by a relative contribution to the ICM enrichment of 80\% -- 100\% for SNCC. Similar results have been reported by \citet{2010PASJ...62.1445S} for the fossil group NGC 1550. Also \citet{2007MNRAS.380.1554R} found that for their galaxy group sample outside the cool core SNCC dominate the enrichment over SNIa. 
Recent results from \citet{2015arXiv150606164S} from Suzaku observations of the Virgo Cluster gave comparable results with a relative contribution of $79\%-85\%$ for SNCC indicating a similar enrichment history for galaxy groups and clusters. They measured abundance ratios beyond the virial radius for the first time and ruled out an enrichment of solely SNCC at large radii at 9$\sigma$ level. However, the authors note that owing to the limited accuracy of the SN yield models uncertainties in the measurements still remain.

\renewcommand{\arraystretch}{1.2}
\begin{table}
\caption{Ratio of the relative number of Supernovae Type II for the elements Si and S and two different SNIa yield models.}             
\centering                                      
\begin{tabular}{l c c}          
\hline \hline                      
 &$ R_{\rm Si}  $ & $  R_{\rm S}   $    \\
W7 &$ 0.81 _{ -0.15 }^{+ 0.14 }$&$ > 0.91 $\\
WDD2 &$ 0.80 _{ -0.18 }^{+ 0.15 }$&$ > 0.90 $\\
\hline \hline                                             
\end{tabular}
\label{tab:sn_ratio}
\end{table}
\renewcommand{\arraystretch}{1}

\subsection{Gas mass and total mass}\label{sec:discussion_mass}
With the deprojected gas density and the temperature profiles we computed the X-ray hydrostatic mass of the galaxy group using the hydrostatic equation

\begin{equation}\label{eq:mtot}
 M_{\rm tot}(<R) = -\frac{kT_{\rm gas}R}{G\mu m_{\rm p}}\left(\frac{{\rm d\,ln\,}\rho_{\rm gas}}{{\rm d\,ln\,}R}+\frac{{\rm d\,ln\,}T_{\rm gas}}{{\rm d\,ln\,}R}\right)
\end{equation}
 with $m_{\rm p}$ being the proton mass, $\mu$ the mean molecular weight, and G the gravitational constant. We find a value of $M(<R_{\rm 200}) = (1.02^{+0.04}_{-0.04})\times10^{14}$\,M$_{\odot}$. 
Using the density profile we obtain an estimate for $R_{200}$ yielding $R_{200} = 23.7\arcmin$. This value is considerably lower than the first estimate from the RASS data.

\begin{figure}
\resizebox{\hsize}{!}{\includegraphics{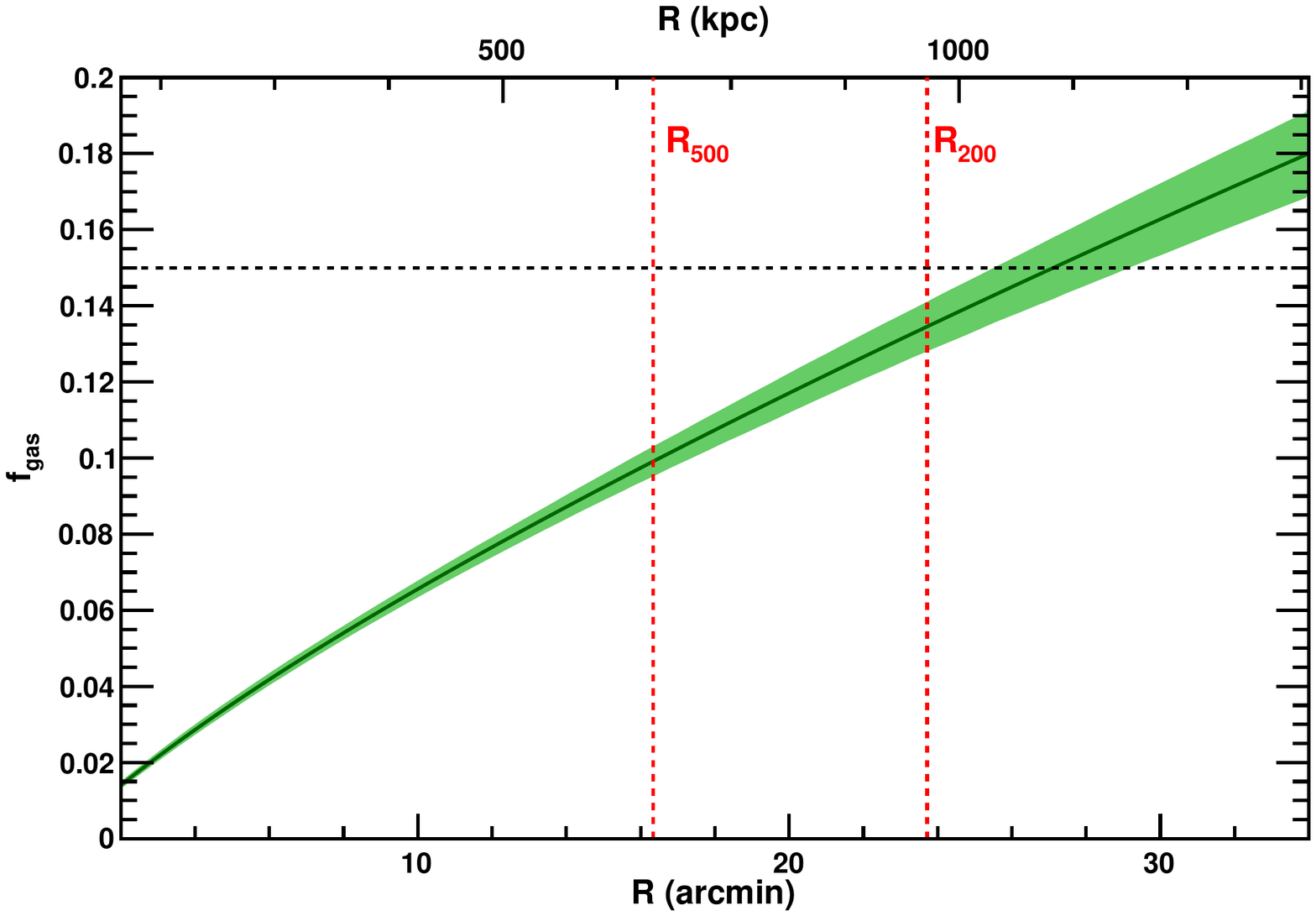}}
\caption{Gas mass fraction profile of UGC03957. The horizontal dashed line shows the cosmic mean value of 0.15 (\citealp{2014A&A...571A..16P}). The shaded area corresponds to the 68\% confidence region.}
\label{fig:fgas}
\end{figure}

We obtain the gas mass fraction profile as shown in Fig. \ref{fig:fgas} (the innermost part is not shown owing to our simplified temperature model that does not describe the cool core). Up to $R_{500}$ the gas mass fraction is below 10\%, which is a typical value found for galaxy groups as in, e.g., \citet{2015A&A...573A.118L}, \citet{2009ApJ...693.1142S} and \citet{2012ApJ...748...11H}. Galaxy clusters typically show somewhat higher gas mass fraction above $0.1$ as found by, e.g., \citet{2006ApJ...640..691V}. In galaxy groups feedback processes have more effect than in clusters and lead to further expulsion of the gas. Beyond $R_{500}$ the fraction of UGC03957 rises up to ${\sim} 13\%$ at $R_{200}$ and ${\sim}18\%$ at the maximum radius we reach with our observation, which is slightly above the cosmic mean value. This behavior is in contrast to measurements by, e.g., \cite{2011Sci...331.1576S} for the Perseus cluster and \citet{2012MNRAS.424.1826W} for the cluster PKS 0745−191. They measured gas mass fractions of ${\sim}0.23$ and ${\sim}0.19$ already around $R_{200}$, respectively, while UGC03957 only rises above this value far beyond $R_{200}$. \citet{2013A&A...551A..23E} investigated the gas properties for a sample of 18 galaxy clusters with combined ROSAT and Planck data. They found $f_{\rm gas}$ around 18\% beyond $R_{200}$ in agreement with our findings. A likely explanation for this excess is a deviation from hydrostatic equilibrium. If the assumption of hydrostatic equilibrium is violated this can result in a lower total mass estimate and therefore a higher gas mass fraction. \cite{2008A&A...491...71P} showed -- by performing N-body/SPH simulations of about 100 galaxy clusters -- that masses can be underestimated by up to $15\%$ at $R_{200}$.

\subsection{Entropy profile}\label{sec:discussion_entropy}

\begin{figure}
\resizebox{\hsize}{!}{\includegraphics{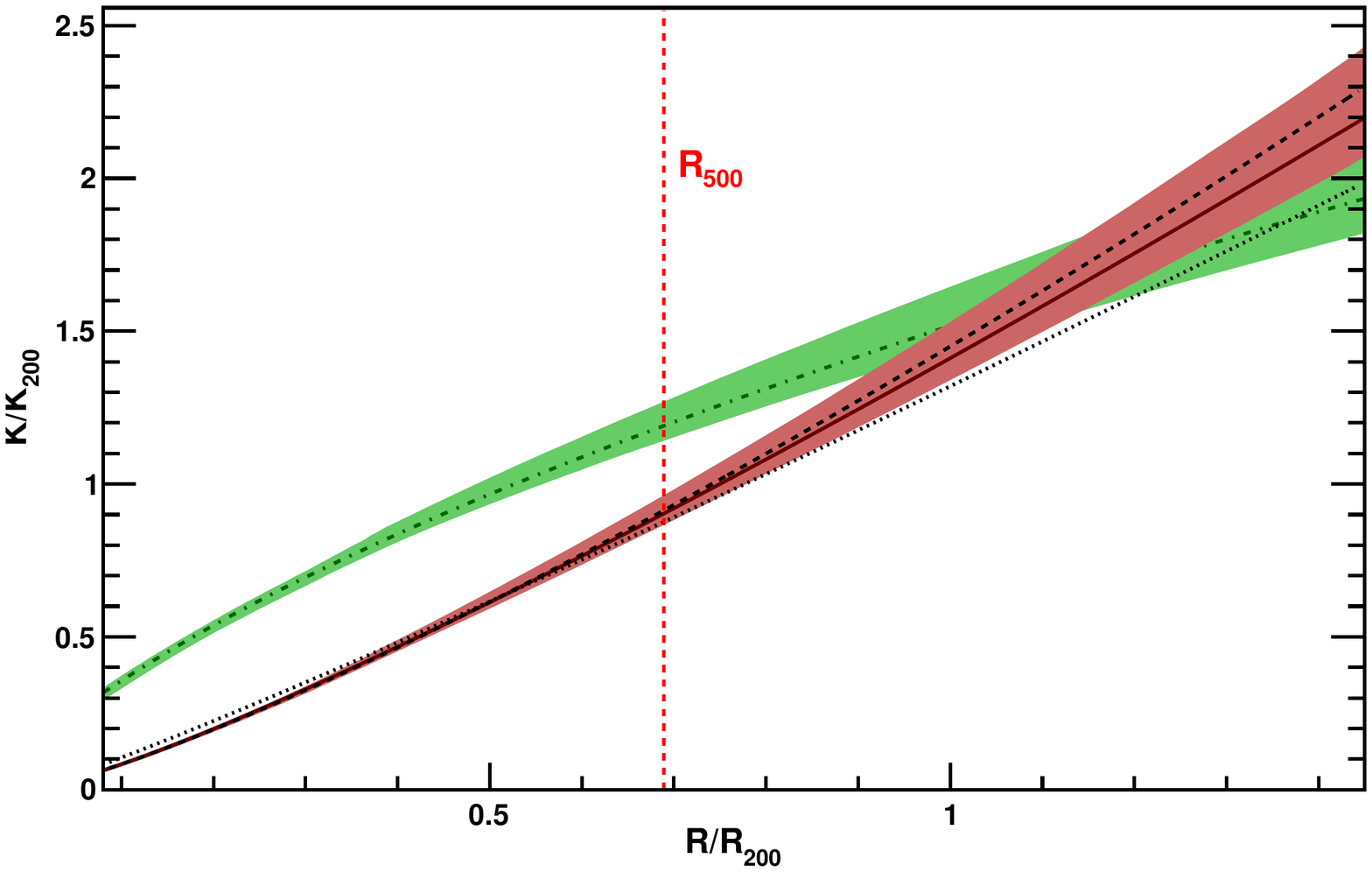}}
\caption{Scaled entropy profile of UGC03957. The green dot-dashed line shows the entropy profile estimated from the deprojected temperature and density profiles. The red solid line represents the profile corrected for the gas mass fraction as suggested by \citet{2010A&A...511A..85P} (Eq. \ref{eq:K_corr}). Dotted and dashed lines represent the predictions from gravitational structure formation simulations by \citet{2005RvMP...77..207V} and correspond to Eq. \ref{eq:voit} and Eq. \ref{eq:voit_freeslope}, respectively.}
\label{fig:entropy_profile}
\end{figure}

A good indicator for the hydrodynamical status of the ICM is the entropy, which we obtained by combining the deprojected density and temperature profiles using the entropy definition $K = {\rm k}Tn_{\rm e}^{-\frac{2}{3}}$ with Boltzmann constant k.
The derived profile is shown in Fig. \ref{fig:entropy_profile} in green. \citet{2005RvMP...77..207V} performed non-radiative simulations from gravitational structure formation and found that their simulated cluster sample follows the relation 
\begin{equation}\label{eq:voit}
 \frac{K_{\rm sim}}{K_{\rm 200}} = 1.32\left(\frac{R}{R_{\rm 200}}\right)^{1.1}
\end{equation}
 with the normalization

 \begin{equation}
 {K_{\rm 200}} = 362\frac{GM_{\rm 200}\mu m_{\rm p}}{2R_{200}}\left(\frac{1}{{\rm keV}}\right)\times\left(\frac{H(z)}{H_0}\right)^{-\frac{4}{3}}\left(\frac{\Omega_m}{0.3}\right)^{-\frac{4}{3}}\,{\rm keV\,cm^{-2}}.
\end{equation}

However, in this fit to the simulated data the slope was fixed to the common literature value of 1.1, but the authors note that outside $0.2R_{200}$ their sample seems to indicate a slightly steeper slope. For this reason they performed another fit with free slope and normalization and found 

\begin{equation}\label{eq:voit_freeslope}
 \frac{K_{\rm sim}}{K_{\rm 200}} = 1.45\left(\frac{R}{R_{\rm 200}}\right)^{1.24}.
\end{equation}

Fig. \ref{fig:entropy_profile} shows both fits together with our measurement (green line and shaded area corresponding to the 68\% confidence region). 
At ${\sim} R_{200}$ our measurements agree with the expectation, but at smaller radii we find a clear entropy excess compared to the numerical prediction. \citet{2010A&A...511A..85P} analyzed 31 nearby clusters and found a similar behavior for their sample, i.e., many entropy profiles showing larger deviation towards the central regions. They reported that the profiles match well with the numerical simulations by \citet{2005RvMP...77..207V} when a gas mass fraction correction is applied, which also reduces the scatter in the entropy profiles significantly. The correction is as follows,

\begin{equation}\label{eq:K_corr}
 K_{\rm corr} = K_{\rm measure}\times f_{\rm gas}(<R)^{2/3}f_{\rm b}^{-2/3},
\end{equation}
with the cosmic baryon fraction $f_b = 0.15$ (\citealp{2013A&A...550A.134P}). We applied the correction to our entropy profile (in red in Fig. \ref{fig:entropy_profile}). The resulting profile is in  much better agreement with Eq. \ref{eq:voit} with a fixed slope of 1.1. Compared to Eq. \ref{eq:voit_freeslope} (the fit to the simulated cluster sample of \citet{2005RvMP...77..207V} with free slope) we even find a perfect agreement. As in their sample, our measurement for UGC03957 suggests a slightly steeper slope than the literature value of 1.1. \citet{2010A&A...511A..85P} discussed several possible explanations for entropy modification. Pre-heating processes or AGN feedback can lift the entropy in the central region as discussed by, e.g., \cite{2010RAA....10.1013W}. Feedback from the central AGN or convection and bulk motion can push the central gas farther outwards or even eject gas from the object, especially in low mass systems with a shallower gravitational potential well, leading to higher entropy. \citet{2010RAA....10.1013W} measured entropy profiles for 31 galaxy groups and clusters and found a clear central entropy excess for all objects. They compared their observation with observationally constrained supernovae explosion rates and also the contribution of AGN feedback and concluded that AGNs can be responsible for the excess entropy. However, the observations were performed with the Chandra satellite and in most cases only reach $R_{500}$. To explain the excess in our analysis, the described effects must have an impact on the gas out to large radii, which is more probable for the low mass systems. \citet{2010MNRAS.406..822M} explicitly focused on simulations of AGN feedback in galaxy groups. Their simulations reproduce the observations up to $R_{500}$ and the central entropy excess very well.

\citet{2012ApJ...748...11H} and \citet{2013ApJ...775...89S} studied the galaxy groups RXJ1159+5531 and ESO 3060170, respectively, and obtained similar results to our study. Both found an entropy excess out to large radii for their objects and good overall agreement with the simulations by \citet{2005RvMP...77..207V} after applying the $f_{\rm gas}$ correction. {Recently \citet{2015ApJ...805..104S} extended the study of RXJ1159+5531 to full azimuthal completeness and confirmed this entropy behavior.} Nevertheless, \citet{2013ApJ...775...89S} reported an entropy drop at ${\sim} R_{200}$, which is consistent with observations for several galaxy clusters. \citet{2013MNRAS.432..554W} compared the entropy profile for seven clusters obtained in previous analyses to the baseline prediction of \citet{2005RvMP...77..207V} and found a significant {entropy flattening (or even a drop for some clusters)} at large radii. They suggest clumping as one possible explanation, but deviations from thermal equilibrium between electrons and protons can also lower the entropy. However, the results of \citet{2012ApJ...748...11H}, {\citet{2015ApJ...805..104S}}, and also our results closely match the simulation in the outskirts. 
This seems to indicate a difference between galaxy groups and clusters regarding the impact of non-gravitational effects on the entropy profile in the outskirts. 
{We note that using simple models for temperature and density profiles (i.e., power law and single-beta model, respectively) results in a power-law description of the entropy profile that in principle cannot reflect more complicated behavior. However, these models describe our data very well and an entropy flattening is absent for our object. Additionally, we tested a double-beta model for the deprojection and find a consistent profile compared to the single-beta case in the outskirts (cf. Sec. \ref{sec:deprojection}). Hence, we are confident that our models yield a reliable entropy profile.}

However, even for galaxy clusters the entropy measurements are contradictory as shown by \cite{2013A&A...551A..22E} who analyzed 18 clusters and found a better agreement of the average entropy profile with numerical simulations at $R>R_{500}$ in contrast to \citet{2013MNRAS.432..554W}. One point discussed in \cite{2013A&A...551A..22E} is the missing azimuthal completeness in most Suzaku studies, which might explain the different findings. Our study and that of {\citet{2015ApJ...805..104S}} have good azimuthal coverage, while \citet{2013ApJ...775...89S} only studied one azimuthal direction. Thus, good azimuthal coverage may be important in order to obtain overall cluster and group entropy profiles. Nevertheless, to constrain a ``universal entropy profile'' for galaxy clusters and in particular for galaxy groups, larger samples are needed, which then also allow the impact of non-gravitational effects to be studied in greater detail.

\section{Conclusion}\label{sec:conclusion}
We analyzed five Suzaku observations of a galaxy group, reaching ${\sim} 1.4 R_{\rm 200}$. We found that the group is azimuthally symmetric and performed a simultaneous fit of all outskirts observations and determined the temperature, abundance, surface brightness, density, entropy, and $f_{\rm gas}$ profiles. Our main findings are the following:

\begin{itemize}
 \item The surface brightness profile is consistent with previous measurements of galaxy groups with a single-beta model fit to the Suzaku data yielding $\beta=0.55\pm0.01$. 
 Extrapolation of the XMM-Newton best fit double-beta model leads to large deviation from our Suzaku measurements and emphasizes the importance of accurate measurements out to large radii to avoid biases.
 \item The temperature profile drops by almost a factor of three from the center to the outskirts. This is consistent with previous Suzaku analyses for galaxy clusters as shown in \citet{2013SSRv..tmp...33R} and supports a self-similar picture.
 \item The abundance profile shows a flat behavior outside the center which points to galactic winds as the primary enrichment process, as found in simulations by \citet{2007A&A...466..813K}.
 \item The abundance pattern of the group can be explained by a relative contribution of ${\sim}80\%$ -- $100\%$ for SNCC to the ICM enrichment. This is comparable with the results from previous measurements for galaxy groups (e.g., \citealp{2010PASJ...62.1445S} and \citealp{2007ApJ...667L..41S}) and also with recent results for the Virgo cluster (\citealp{2015arXiv150606164S}).
 
 \item The gas mass fraction increases with radius and is ${<}0.1$ inside $R_{500}$ which is comparable to previous results for galaxy groups (e.g., \citealp{2015A&A...573A.118L}, \citealp{2009ApJ...693.1142S}, \citealp{2012ApJ...748...11H}), but is in contrast to galaxy clusters which show higher gas mass fractions. One explanation are feedback processes that act more efficient in galaxy groups. Outside $R_{200}$ the gas mass fraction exceeds the cosmic mean value while inside this radius it stays below ${\sim}13\%$ in contrast to what was previously observed for the Perseus cluster (\citealp{2011Sci...331.1576S}) which showed high gas mass fractions already around $R_{200}$. A likely explanation is the breakdown of hydrostatic equilibrium in the outer parts of the group where the gas has not yet virialized.
 
 \item The measured entropy profile shows an excess compared to numerical simulations performed by \citet{2005RvMP...77..207V}. Correcting the entropy profile as suggested by \citet{2010A&A...511A..85P} leads to good agreement with the simulations and indicates a slightly steeper slope than the expected value of 1.1. We conclude that feedback processes and the redistribution of material, for example due to AGN activity -- especially imprinting in galaxy groups with a lower potential well --, play a major role out to larger radii than in galaxy clusters. Pre-heating processes might also be responsible for the observed excess. Our findings are in agreement with \cite{2013A&A...551A..22E} and the study of a fossil group performed by \citet{2012ApJ...748...11H}, but are in contrast to results for galaxy clusters measured by \citet{2013MNRAS.432..554W}, which showed an entropy drop around $R_{200}$ pointing to non-gravitational effects such as gas clumping or non-equilibrium states in the outskirts of the clusters. This hints at a possible difference between groups and clusters however, azimuthal completeness of the studies is an important factor and might at least partially explain the different findings.
\end{itemize}

\begin{acknowledgements}
We would like to thank Gerrit Schellenberger, Brenda Selene Miranda Ocejo, Bharadwaj Vijaysarathy, Hiroki Akamatsu, Sandra Martin, and Dominik Klaes for fruitful discussions and helpful support. {We also thank the anonymous referee for the helpful comments. S.T. acknowledges support from the DFG through grant RE 1462/6 and Transregional Collaborative Research Centre TRR33 “The Dark Universe” {and the Bonn-Cologne Graduate School of Physics and Astronomy}. L.L. acknowledges support by the DFG through grant LO 2009/1-1 and TRR33}. T.H.R. acknowledges support from the DFG through Heisenberg grant RE 1462/5 and grant RE 1462/6, and from TRR 33. 
\end{acknowledgements}

\bibliographystyle{aa} 
\bibliography{bibtex.bib}
\FloatBarrier
\newpage

\begin{appendix}
\onecolumn
\section{Fit results and systematic uncertainties}
\renewcommand{\arraystretch}{1.2}
\begin{table*}[h]
\caption{Fit results and systematics for the central observation. The CXB in Cols. 3 and 4 was scaled according to Tab. \ref{tab:cxb_fluct}. The NXB was scaled by $\pm 3\%$. Upper limits are given at 90\% confidence level.}              
\label{tab:results_center}     
\centering                                     
\begin{tabular}{c c c c c c c p{3cm}}          
Annulus & Nominal &  CXB $\downarrow$ &CXB $\uparrow$&NXB $\downarrow$& NXB $\uparrow$& Abundance table  \\
&&&&&& \citet{1989GeCoA..53..197A}\\ \hline \hline
&\multicolumn{6}{c}{T (keV)}\\\hline
1&$2.61_{-0.04}^{+0.04}$&$2.61_{-0.04}^{+0.04}$&$2.61_{-0.04}^{+0.04}$&$2.61_{-0.04}^{+0.04}$&$2.61_{-0.04}^{+0.04}$&$2.58_{-0.04}^{+0.04}$\\
2&$3.13_{-0.17}^{+0.16}$&$3.15_{-0.16}^{+0.16}$&$3.11_{-0.17}^{+0.16}$&$3.14_{-0.17}^{+0.16}$&$3.13_{-0.17}^{+0.16}$&$3.03_{-0.17}^{+0.17}$\\
3&$2.43_{-0.18}^{+0.17}$&$2.47_{-0.16}^{+0.15}$&$2.35_{-0.17}^{+0.16}$&$2.45_{-0.18}^{+0.17}$&$2.40_{-0.18}^{+0.17}$&$2.36_{-0.18}^{+0.17}$\\
4&$2.08_{-0.24}^{+0.27}$&$2.12_{-0.19}^{+0.29}$&$1.95_{-0.25}^{+0.20}$&$2.11_{-0.23}^{+0.33}$&$2.04_{-0.27}^{+0.24}$&$2.05_{-0.25}^{+0.24}$\\
\hline
&\multicolumn{6}{c}{Mg}\\\hline
1&$1.02_{-0.22}^{+0.22}$&$1.02_{-0.22}^{+0.22}$&$1.02_{-0.22}^{+0.21}$&$1.02_{-0.22}^{+0.22}$&$1.02_{-0.22}^{+0.22}$&$1.07_{-0.24}^{+0.23}$\\
2&$<0.80$&$<0.82$&$<0.73$&$<0.80$&$<0.80$&$<0.76$\\
\hline
&\multicolumn{6}{c}{Si}\\\hline
1&$1.08_{-0.15}^{+0.16}$&$1.08_{-0.15}^{+0.16}$&$1.07_{-0.15}^{+0.16}$&$1.08_{-0.15}^{+0.16}$&$1.08_{-0.15}^{+0.16}$&$0.98_{-0.15}^{+0.15}$\\
2&$1.05_{-0.42}^{+0.43}$&$1.05_{-0.43}^{+0.43}$&$1.05_{-0.42}^{+0.42}$&$1.06_{-0.42}^{+0.43}$&$1.05_{-0.42}^{+0.43}$&$0.94_{-0.39}^{+0.39}$\\
3&$<0.44$&$<0.46$&$<0.38$&$<0.44$&$<0.44$&$<0.36$\\
\hline
&\multicolumn{6}{c}{S}\\\hline
1&$0.83_{-0.21}^{+0.21}$&$0.84_{-0.21}^{+0.21}$&$0.83_{-0.21}^{+0.21}$&$0.84_{-0.21}^{+0.21}$&$0.83_{-0.21}^{+0.21}$&$0.66_{-0.18}^{+0.18}$\\
2&$1.13_{-0.61}^{+0.61}$&$1.12_{-0.62}^{+0.62}$&$1.14_{-0.61}^{+0.61}$&$1.13_{-0.62}^{+0.61}$&$1.14_{-0.61}^{+0.61}$&$0.92_{-0.50}^{+0.50}$\\
3&$0.57_{-0.45}^{+0.45}$&$0.59_{-0.46}^{+0.46}$&$0.51_{-0.43}^{+0.43}$&$0.57_{-0.46}^{+0.46}$&$0.57_{-0.45}^{+0.45}$&$0.40_{-0.37}^{+0.37}$\\
\hline
&\multicolumn{6}{c}{Ar}\\\hline
1&$2.40_{-0.59}^{+0.59}$&$2.40_{-0.59}^{+0.59}$&$2.40_{-0.59}^{+0.59}$&$2.40_{-0.59}^{+0.59}$&$2.41_{-0.59}^{+0.59}$&$1.60_{-0.43}^{+0.43}$\\
\hline
&\multicolumn{6}{c}{Ca}\\\hline
1&$1.21_{-0.49}^{+0.50}$&$1.21_{-0.49}^{+0.50}$&$1.21_{-0.49}^{+0.50}$&$1.20_{-0.49}^{+0.50}$&$1.21_{-0.49}^{+0.50}$&$1.14_{-0.50}^{+0.50}$\\
\hline
&\multicolumn{6}{c}{Fe}\\\hline
1&$0.92_{-0.06}^{+0.06}$&$0.92_{-0.06}^{+0.06}$&$0.91_{-0.06}^{+0.06}$&$0.92_{-0.06}^{+0.06}$&$0.92_{-0.06}^{+0.06}$&$0.66_{-0.04}^{+0.04}$\\
2&$0.44_{-0.13}^{+0.13}$&$0.44_{-0.13}^{+0.13}$&$0.43_{-0.13}^{+0.13}$&$0.44_{-0.13}^{+0.13}$&$0.43_{-0.13}^{+0.13}$&$0.30_{-0.09}^{+0.10}$\\
3&$0.34_{-0.10}^{+0.11}$&$0.35_{-0.10}^{+0.11}$&$0.31_{-0.09}^{+0.11}$&$0.35_{-0.10}^{+0.11}$&$0.33_{-0.10}^{+0.11}$&$0.23_{-0.07}^{+0.08}$\\
4&$0.43_{-0.17}^{+0.20}$&$0.45_{-0.17}^{+0.21}$&$0.35_{-0.15}^{+0.19}$&$0.44_{-0.17}^{+0.22}$&$0.42_{-0.17}^{+0.20}$&$0.31_{-0.12}^{+0.14}$\\
\hline
&\multicolumn{6}{c}{Norm$^*$}\\\hline
1&$2.30_{-0.04}^{+0.04}$&$2.30_{-0.04}^{+0.04}$&$2.30_{-0.04}^{+0.04}$&$2.30_{-0.04}^{+0.04}$&$2.30_{-0.04}^{+0.04}$&$2.18_{-0.03}^{+0.03}$\\
2&$2.14_{-0.04}^{+0.04}$&$2.15_{-0.04}^{+0.04}$&$2.14_{-0.04}^{+0.04}$&$2.14_{-0.04}^{+0.04}$&$2.14_{-0.04}^{+0.04}$&$2.06_{-0.04}^{+0.04}$\\
3&$2.02_{-0.05}^{+0.05}$&$2.02_{-0.05}^{+0.05}$&$2.04_{-0.05}^{+0.05}$&$2.02_{-0.05}^{+0.05}$&$2.02_{-0.05}^{+0.05}$&$1.96_{-0.05}^{+0.05}$\\
4&$1.67_{-0.08}^{+0.08}$&$1.69_{-0.08}^{+0.07}$&$1.70_{-0.08}^{+0.09}$&$1.68_{-0.08}^{+0.08}$&$1.67_{-0.08}^{+0.08}$&$1.61_{-0.08}^{+0.08}$\\
\hline\hline
&\multicolumn{6}{c}{XRBG}\\\hline
norm$_{\rm CXB}^\dagger$&$1.22_{-0.10}^{+0.10}$&fix&fix&$1.23_{-0.10}^{+0.10}$&$1.21_{-0.10}^{+0.10}$&$1.21_{-0.10}^{+0.10}$\\
norm$_{\rm MWH}^{\circ}$&$4.30_{-1.42}^{+2.82}$&$4.89_{-1.08}^{+2.16}$&$1.88_{-1.09}^{+2.18}$&$4.23_{-1.41}^{+2.86}$&$4.37_{-1.44}^{+2.85}$&$3.46_{-0.98}^{+1.94}$\\
$T_{\rm LHB}$ ($10^{-2}$\,keV)&$9.87_{-0.47}^{+0.91}$&$9.87_{-0.47}^{+0.91}$&$9.87_{-0.47}^{+0.91}$&$9.87_{-0.47}^{+0.91}$&$9.87_{-0.47}^{+0.91}$&$10.06_{-0.46}^{+0.90}$\\
norm$_{\rm LHB}^{\circ}$&$9.80_{-0.43}^{+0.86}$&$9.81_{-0.43}^{+0.86}$&$9.78_{-0.43}^{+0.86}$&$9.80_{-0.43}^{+0.86}$&$9.81_{-0.43}^{+0.86}$&$8.70_{-0.36}^{+0.72}$\\

\hline \hline                                             
\end{tabular}
\begin{minipage}{1.8\columnwidth}
\vspace{0.2cm}
\footnotesize $^*$ ${\rm norm}=\frac{1}{4\pi[D_A(1+z)]^2}\int n_{\rm e}n_{\rm H}{\rm d}V\,10^{-16}\,{\rm cm}^{-5}$ with $D_A$ being the angular diameter distance to the source.\\
$^{\circ}$ Normalization of the apec component scaled to area 400$\pi$ assumed in the uniform-sky ARF calculation. \\ \phantom{-,}norm = $\frac{1}{4\pi[D_A(1+z)]^2}\int n_{\rm e}n_{\rm H}{\rm d}V$\,10$^{-20}$\,cm$^{-5}$ .
\end{minipage}
\begin{minipage}{1.8\columnwidth}
\vspace{0.03cm}
\footnotesize  $^\dagger$ in units of $10^{-3}$ photons/s/cm$^2$ at 1\,keV scaled to the area 400$\pi$.
 \end{minipage}
\end{table*}
\renewcommand{\arraystretch}{1}

\newpage

\renewcommand{\arraystretch}{1.2}
\begin{table*}
\caption{Fit results and systematics for the simultaneous fit of the north, east, south, and west observations. The CXB in Cols. 3 and 4 was scaled according to Tab. 3. The NXB was
scaled by $\pm$3\%.}              
\label{tab:results_outskirts}      
\centering                                      
\begin{tabular}{c c c c c c c}          
Annulus & Nominal &  CXB $\downarrow$ &CXB $\uparrow$&NXB $\downarrow$& NXB $\uparrow$& Abundance table \\
&&&&&&\citet{1989GeCoA..53..197A} \\ \hline \hline
&\multicolumn{6}{c}{T (keV)}\\\hline

5&$1.20_{-0.10}^{+0.07}$&$1.22_{-0.09}^{+0.07}$&$1.18_{-0.11}^{+0.08}$&$1.19_{-0.10}^{+0.07}$&$1.20_{-0.10}^{+0.07}$&$1.20_{-0.10}^{+0.07}$\\
6&$1.18_{-0.09}^{+0.07}$&$1.21_{-0.08}^{+0.07}$&$1.17_{-0.10}^{+0.07}$&$1.18_{-0.09}^{+0.07}$&$1.19_{-0.09}^{+0.07}$&$1.18_{-0.09}^{+0.07}$\\

\hline
&\multicolumn{6}{c}{Z ($Z_\odot$)}\\\hline
5&$0.28_{-0.11}^{+0.17}$&$0.24_{-0.08}^{+0.11}$&$0.35_{-0.15}^{+0.24}$&$0.31_{-0.12}^{+0.20}$&$0.26_{-0.10}^{+0.15}$&$0.20_{-0.07}^{+0.11}$\\
6&$0.39_{-0.15}^{+0.29}$&$0.31_{-0.10}^{+0.15}$&$0.54_{-0.23}^{+0.47}$&$0.45_{-0.18}^{+0.40}$&$0.35_{-0.13}^{+0.23}$&$0.25_{-0.09}^{+0.15}$\\

\hline
&\multicolumn{6}{c}{Norm$^*$}\\\hline

5&$1.48_{-0.39}^{+0.43}$&$1.74_{-0.33}^{+0.38}$&$1.20_{-0.34}^{+0.42}$&$1.37_{-0.39}^{+0.43}$&$1.59_{-0.39}^{+0.44}$&$1.58_{-0.37}^{+0.41}$\\
6&$1.84_{-0.64}^{+0.67}$&$2.30_{-0.49}^{+0.55}$&$1.35_{-0.51}^{+0.58}$&$1.61_{-0.64}^{+0.67}$&$2.06_{-0.64}^{+0.67}$&$2.12_{-0.61}^{+0.64}$\\

\hline
&\multicolumn{6}{c}{XRBG}\\\hline
norm$_{\rm CXB}^\dagger$&$1.24_{-0.03}^{+0.03}$&fix&fix&$1.27_{-0.03}^{+0.03}$&$1.22_{-0.03}^{+0.03}$&$1.24_{-0.03}^{+0.03}$\\
norm$_{\rm MWH}^{\circ}$&$6.32_{-0.96}^{+1.91}$&$6.52_{-0.94}^{+1.88}$&$6.14_{-0.95}^{+1.90}$&$6.12_{-0.95}^{+1.90}$&$6.51_{-0.96}^{+1.92}$&$4.56_{-0.67}^{+1.34}$\\
$T_{\rm LHB}$ ($10^{-2}$\,keV)&$9.71_{-0.48}^{+0.93}$&$9.73_{-0.47}^{+0.93}$&$9.70_{-0.48}^{+0.93}$&$9.71_{-0.48}^{+0.93}$&$9.72_{-0.48}^{+0.93}$&$9.91_{-0.47}^{+0.92}$\\
norm$_{\rm LHB}^{\circ}$&$9.65_{-0.43}^{+0.85}$&$9.67_{-0.43}^{+0.85}$&$9.63_{-0.43}^{+0.85}$&$9.64_{-0.43}^{+0.85}$&$9.66_{-0.43}^{+0.85}$&$8.60_{-0.36}^{+0.71}$\\

\hline \hline                                            
\end{tabular}
\begin{minipage}{1\columnwidth}
\vspace{0.2cm}
\footnotesize $^*$ ${\rm norm}=\frac{1}{4\pi[D_A(1+z)]^2}\int n_{\rm e}n_{\rm H}{\rm d}V\,10^{-16}\,{\rm cm}^{-5}$ with $D_A$ being the angular diameter distance to the source and rescaled to the central observation for better comparability.\\
$^{\circ}$ Normalization of the apec component scaled to area 400$\pi$ assumed in the uniform-sky ARF calculation. \\ \phantom{-,}norm = $\frac{1}{4\pi[D_A(1+z)]^2}\int n_{\rm e}n_{\rm H}{\rm d}V$\,10$^{-20}$\,cm$^{-5}$ .
\end{minipage}
\begin{minipage}{1\columnwidth}
\vspace{0.03cm}
\footnotesize  $^\dagger$ in units of $10^{-3}$ photons/s/cm$^2$ at 1\,keV scaled to the area 400$\pi$.
 \end{minipage}
\end{table*}
\renewcommand{\arraystretch}{1}

\end{appendix}
\end{document}